\documentclass
[aps,twocolumn,pra,superscriptaddress,amsmath,showpacs,tightenlines,twoside]{revtex4}%
\usepackage{amsfonts}
\usepackage{amsmath}
\usepackage{amssymb}
\usepackage{version}
\usepackage[truetex]{graphicx}
\usepackage[truetex]{color}
\usepackage{float}
\usepackage{hyperref}%
\providecommand{\U}[1]{\protect\rule{.1in}{.1in}}
\makeatletter
\newcommand\figcaption{\def\@captype{figure}\caption}
\newcommand\tabcaption{\def\@captype{table}\caption}
\makeatother
\begin{document}
\title{Realization of microwave amplification, attenuation, and frequency conversion
using a single three-level superconducting quantum circuit}
\author{Yan-Jun Zhao}
\affiliation{Institute of Microelectronics, Tsinghua University, Beijing 100084, China}
\affiliation{Beijing National Laboratory for Condensed Matter Physics}
\affiliation{Institute of Physics, Chinese Academy of Sciences, Beijing 100190, China}
\author{Jiang-Hao Ding}
\affiliation{Institute of Microelectronics, Tsinghua University, Beijing 100084, China}
\author{Z. H. Peng}
\affiliation{CEMS, RIKEN, Saitama 351-0198, Japan}
\author{Yu-xi Liu}
\email{yuxiliu@mail.tsinghua.edu.cn}
\affiliation{Institute of Microelectronics, Tsinghua University, Beijing 100084, China}
\affiliation{Tsinghua National Laboratory for Information Science and Technology (TNList),
Beijing 100084, China}
\date{\today }

\begin{abstract}
Using different configurations of applied strong driving and weak probe
fields, we find that only a single three-level superconducting quantum circuit
(SQC) is enough to realize amplification, attenuation and frequency conversion
of microwave fields. Such a three-level SQC has to possess $\Delta$-type
cyclic transitions. Different from the parametric amplification (attenuation)
and frequency conversion in nonlinear optical media, the real energy levels of
the three-level SQC are involved in the energy exchange when these processes
are completed. We quantitatively discuss the effects of amplification
(attenuation) and the frequency conversion for different types of driving
fields. The optimal points are obtained for achieving the maximum
amplification (attenuation) and conversion efficiency. Our study provides a
new method to amplify (attenuate) microwave, realize frequency conversion, and
also lay a foundation for generating single or entangled microwave photon
states using a single three-level SQC.

\end{abstract}

\maketitle

\pagenumbering{arabic}

\section{Introduction}

Three-wave mixing is very fundamental in nonlinear quantum
optics~\cite{NonlinearOptics}. It can be used to generate single photon or
entangled photon pairs. Three-wave mixing can also be used to convert the
frequency of the weak electromagnetic field or amplify the weak
electromagnetic signal by virtue of another strong driving field. In atomic
systems with inversion symmetry, the three-wave mixing can only occur in a
parametric way because of the selection rule in the electric-dipole
interaction between the atoms and electric fields. Thus, the nonlinear
interaction strength of the three-wave mixing is usually weak in the natural
atomic systems. In contrast to the parametric weak nonlinearity realized by
the virtual single- or two-photon processes, the nonlinear interaction
strength can be increased significantly when the real energy exchange is
involved. For example, in molecular
systems~\cite{Three-wave-1977,Molecular-book,Patterson-PRL}, the inversion
symmetry can be broken, and the transitions between any two energy levels are
possible. In such cases, the three-wave mixing processing can be realized
using real energy transitions between energy levels. In artificial atoms,
e.g., semiconducting quantum dots or superconducting quantum circuits (SQCs),
the inversion symmetry of their potential energy can be artificially
controlled by externally applied field, that is, the selection rule of the
artificial atoms can be engineered. Thus, they provide us a new platform to
manipulate and engineer linear and nonlinear quantum processes. For
example,the single photon strong coupling can be achieved between microwave
and mechanical modes using a superconducting flux qubit~\cite{ZYXue-APL}. It
is also possible to realize three-wave mixing using real energy transitions
between energy levels of a single SQC~\cite{liuyx}.

The SQCs~\cite{RMP,wendin,clarke,xiang,you-today,you-nature,Girvin-review} are
extensively explored for the realization of qubits which are basic building
blocks of quantum information processing. These superconducting artificial
atoms can also be used not only to demonstrate phenomena occurred in atomic
physics and quantum optics, but also to show some novel results, which cannot
be found in natural atomic systems. For example, the dressed
states~\cite{liu2006,Cohen-Tannoudjibook} have been experimentally
demonstrated using superconducting charge qubit
circuits~\cite{delsing2007,delsing2010}. The Autler-Townes
splitting~\cite{atsET,atsEP,atsEP1,atsEF,Bsanders,atsET-2,atsET-3} and
coherent population trapping~\cite{CPT} have also been observed in three-level
SQCs. Experimentalists are trying to find a way to realize the
electromagnetically induced transparency in varieties of three-level
superconducting quantum devices~\cite{CPYang,murali,dutton,HouIan,huichen Sun}. Moreover, the coexistence
of single- and two-photon processes~\cite{liuprl} in three-level
superconducting flux qubit circuits~\cite{liuprl,orlando} has been
experimentally demonstrated~\cite{naturephysics2008} by designed circuit QED
systems. However, such phenomenon cannot be demonstrated using three-level
natural atoms because of the electric-dipole transition rule.

The microwave amplification has been experimentally demonstrated by a single
three-level artificial atom in open one-dimensional space~\cite{Olig}, a
dressed~\cite{Ilichev} or a doubly-dressed superconducting
flux~\cite{Ilichev1} qubit circuit. The coherent frequency conversion has be
demonstrated using two internal degrees of freedom of a single dc-SQUID phase
qubit circuit~\cite{Bussion}. The parametric down conversion and squeezing of
two-mode quantized microwave fields using circuit QED are theoretically
studied~\cite{KMoon}. Moreover, the parametric three-wave mixing devices have
also been experimentally demonstrated~\cite{devoret1,devoret2}. It was shown
that the amplifiers with single artificial atoms~\cite{Olig} are different
from the Josephson junction based parametric three-wave mixing
devices~\cite{devoret1,devoret2} or amplifiers~\cite{A1,A2,A3,A4}. A main
difference is that the real energy exchange between discrete energy levels is
involved when the single artificial atom amplifiers are implemented. We
recently showed that a three-level flux qubit circuits~\cite{liuyx} with
$\Delta$-type cyclic transitions can be used to generate three-wave mixing.

Motivated by the work~\cite{Olig,liuyx}, we now give a detailed study on the
microwave amplification and frequency conversion by a single three-level SQC
with $\Delta$-type transitions, which can be realized by superconducting flux
~\cite{orlando,liuprl}, phase~\cite{Martinis,Hakonen1,Hakonen2},
fluxonium\cite{fluxonium1,fluxonium2} or Xmon qubit circuits~\cite{Xmon}. For
concreteness, we here focus on flux qubit circuits. In our study, we mainly
study the conversion efficiency and the amplification and attenuation of the
weak signal field. Our paper is organized as follows. In Sec.~\ref{sec:model},
we give an overview of the theoretical model. In Sec.~\ref{sec:d12}, we study
the microwave attenuation and frequency conversion for driving type (1). In
Sec.~\ref{sec:d23}, we study the same contents for driving type (2). In Sec.~\ref{sec:d13}, we study microwave amplification, attenuation, and frequency conversion for driving type (3). Finally, we summarize
our results and discuss both measurements and experimental feasibilities.

\begin{figure}[ptb]
\includegraphics[bb=37 36 557 514,width=0.48\textwidth, clip]{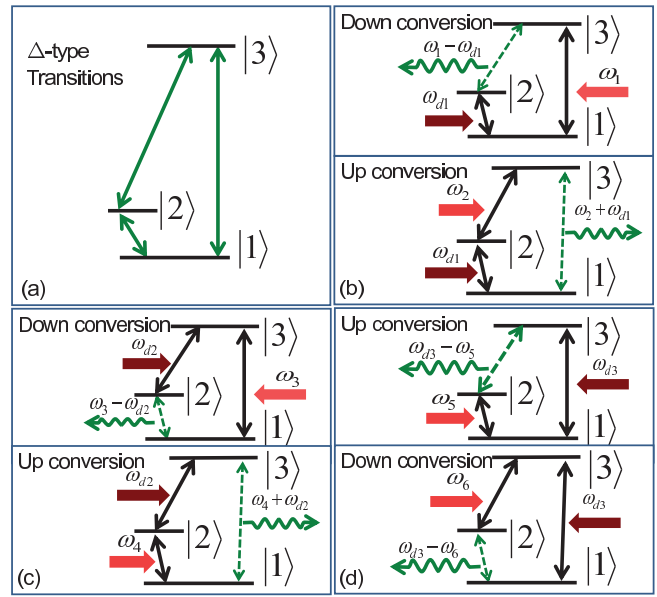}\caption{Schematic
diagram for three-level superconducting quantum circuits with the $\Delta
$-type transitions in (a). The driving field has three different ways to be
applied for \color{blue}attenuation, amplification, or frequency conversion
processes as shown in (b), (c) and (d). \color{black} Each of three figures
includes two panels, corresponding to up and down frequency conversions,
respectively. In all the three figures, the light red arrows mean the signal
field. However, the dark red arrows mean the strong driving field. The green
arrows denote the converted signal.}%
\label{fig1}%
\end{figure}

\section{Theoretical model overview\label{sec:model}}

We study microwave amplifications, attenuation and frequency conversions in a
three-level superconducting artificial atom with $\Delta$-type
transitions~\cite{liuprl}, as schematically shown in Fig.~\ref{fig1}(a). The
three energy levels are denoted by $|1\rangle$, $|2\rangle$, and $|3\rangle$.
Here, we specify our study to superconducting flux qubit circuits, typically consisting
of three Josephson junctions~\cite{orlando,liuprl}. We assume that the bias
magnetic flux is not at the optimal point so that the three energy levels
chosen from such circuit can possess $\Delta$-type transition when the
external fields are applied. We also assume that the three-level system is
placed inside an open one-dimensional transmission line resonator as in
Ref.~\cite{Olig}. Hereafter, for simplicity, we use three-level systems to
denote such three-level superconducting flux qubit circuits. 

To realize the microwave amplification, attenuation, or frequency conversion,
a strong driving field has to be applied to the three-level system with
$\Delta$-type transitions, as shown in Figs.~\ref{fig1}(b), (c) and (d), the
strong driving field can be applied to couple: (1) the energy levels
$|1\rangle$ and $|2\rangle$; or (2) the energy levels $|2\rangle$ and
$|3\rangle$; or (3) the energy levels $|1\rangle$ and $|3\rangle$.
Corresponding to each configuration of the driving field, the signal field can
be applied in two ways. For example, as shown in two panels of Fig.~\ref{fig1}%
(b), the signal field can be applied to either the energy levels $|1\rangle$
and $|3\rangle$ (shown in up panel of Fig.~\ref{fig1}(b)), or the energy
levels $|2\rangle$ and $|3\rangle$ (shown in down panel of Fig.~\ref{fig1}(b))
when the driving field is applied to the energy levels $|1\rangle$ and
$|2\rangle$.

In strong driving types (1) and (2) as shown in Figs.~\ref{fig1}(b) and (c),
the frequency down or up conversion can be realized by properly applying the
signal field. But in the driving type (3) as shown in Fig.~\ref{fig1}(d), the
up and down conversions are a little bit different from the types (1) and (2).
In the types (1) and (2), the down (up) frequency conversion is the difference
between (sum of) frequencies of the signal and driving fields. However, in the
type (3), both the down and up frequency conversions are the difference
between the frequencies of signal and driving fields, when the frequency
difference is larger than the signal frequency, we call this process as the up
conversion, otherwise the down conversion.

\begin{figure}[ptb]
\includegraphics[width=0.48\textwidth,clip]{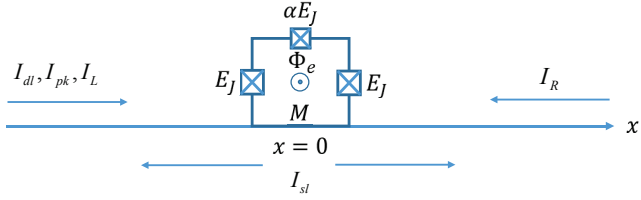}\caption{Schematic
diagram for a typical experimental circuit. The three-junction flux qubit is
coupled to an one-dimensional open transmission line by a edge-sharing mutual
inductance $M$. The qubit is also biased by a DC magnetic flux $\Phi_{e}$. The
location on the transmission line is denoted by $x$ with the qubit placed at
$x=0$. The incident driving and probing currents are respectively denoted by
$I_{dl}\equiv I_{dl}(x,t)$ and $I_{pk}\equiv I_{pk}(x,t)$. The scattered
current is $I_{sl}\equiv I_{sl}(x,t)$. Here, the values of $k=2l-1$ or
$2l$. The quantum noise currents come from both directions are respectively
$I_{L}=I_{L}(x,t)$ and $I_{R}=I_{R}(x,t)$. The Josephson energies
(capacitances) of the two identical junctions and the smaller one are $E_{J}$
($C$) and $\alpha E_{J}$ ($\alpha C$) with $0.5<\alpha<1$, respectively. Here,
$E_{C}=e^{2}/2C$ is called the charging energy with the electron charge $e$.
In addition, the characteristic impedance of the transmission line is assumed
as $Z_{T}$.}%
\label{fig:circuit}%
\end{figure}

In all three driving ways with applied signal fields, the total Hamiltonian
can be generally given by
\begin{equation}
H^{(l)}=\hbar\omega_{21}\sigma_{22}+\hbar\omega_{31}\sigma_{33}+H_{R}%
^{(l)}+H_{pk}^{(l)}+H_{T}^{(l)},\label{eq:H_Total}%
\end{equation}
from Ref.~\cite{liuyx} when one of the weak fields is replaced by a strong
field. We note that we sometimes also call signal fields as probing fields.
Hereafter, $\sigma_{mn}=\left\vert m\right\rangle \left\langle n\right\vert $
with $n$, $m=1,\,2,$ or $3$. The superscript $l=1,\,2$, or $3$ is used to
represent different driving types. The Hamiltonian $H^{(l)}$ can be
exemplified by the circuit schematic in Fig.~\ref{fig:circuit} where the
signals applied through the transmission line will be scattered by the flux
qubit as in Ref.~\cite{nec2010science}. In the type of the $l$th driving, the
symbol $H_{R}^{(l)}\equiv H_{R}^{(l)}(t)$ (or $H_{pk}^{(l)}\equiv H_{pk}%
^{(l)}(t)$) denotes the interaction Hamiltonian between the strong driving
field (or the weak signal field) and the three-level system. The Hamiltonian
\begin{equation}
H_{T}^{(l)}\equiv H_{T}^{(l)}\left(  t\right)  =-M\hat{I}\left(  I_{L}\left(
0,t\right)  +I_{R}\left(  0,t\right)  \right)  \label{eq:H_T_l}%
\end{equation}
describes the interaction between the three-level system and environment in
open one-dimensional transmission line. In Eq.~(\ref{eq:H_T_l}), we have used
the two following symbols
\begin{align}
I_{L}\left(  x,t\right)   &  =-\int_{-\infty}^{\infty}\text{d}\omega g\left(
\omega\right)  A_{L}\left(  \omega\right)  e^{-i\omega\left(  t-x/v\right)
},\label{eq:I_L}\\
I_{R}\left(  x,t\right)   &  =\int_{-\infty}^{\infty}\text{d}\omega g\left(
\omega\right)  A_{R}\left(  \omega\right)  e^{-i\omega\left(  t+x/v\right)
},\label{eq:I_R}%
\end{align}
with $g\left(  \omega\right)  =i\operatorname{sgn}\left(  \omega\right)
\sqrt{\hbar\left\vert \omega\right\vert /4\pi Z_{T}}$ for the noise currents
coming from the left and right. The parameter $Z_T$ represents the characteristic impedence of the transmission line. The commutation relations of $A_{\alpha
}\left(  \omega\right)  $ are $\left[  A_{\alpha}\left(  \omega_{1}\right)
,A_{\alpha}\left(  \omega_{2}\right)  \right]  =\delta\left(  \omega
_{1}+\omega_{2}\right)  \operatorname{sgn}\left(  \omega_{1}\right)  $ for
$\alpha=L$ or $R$. Without the driving fields, the relaxation rates of the three-level system are proportional to the parameters
\begin{equation}
\lambda_{mn}=I_{mn}^{2}\omega_{mn},m>n.
\end{equation}
All the types of interaction Hamiltonians $H_{R}^{(l)}(t)$ and $H_{pk}%
^{(l)}(t)$ have been summarized in Table~\ref{tab1}.

Our research is based on the following hypotheses. (1) The intrinsic loss of
the three-level system is negligible. Hence, the 1D open space determines the
total decay rates. (2) The frequency shifts induced by the driving field are
much larger than the decay rates but still negligibly small compared to the
original eigen frequencies of the three-level system. (3) The interaction
Hamiltonian between the flux qubit circuit and the probe field $H_{pk}%
^{\left(  l\right)  }\left(  t\right)  $ is a small quantity compared with
that between the flux qubit circuit and the driving field $H_{R}^{\left(
l\right)  }\left(  t\right)  $. Thus, the response to the probe signal can be
solved using linear response theory~\cite{NonlinearOptics}. (4) The
environment temperature is too enough to induce effective dephasing or thermal excitation.

Below we will first use driving type (1) as an example to show detailed
derivations. The treatments in driving type (2) and (3) are similar to that in
driving type (1).

\begin{widetext}
\begingroup
%\squeezetable
\begin{table*}
\caption{\label{tab1} Summary of the interaction Hamiltonian $H_{R}^{(l)}(t)$ between
the three-level superconducting artificial atoms and the strong driving field,
which can be applied to the three-level systems in three different ways.
Corresponding to each strong driving field, the probe fields can be applied to the
three-level system in two different ways with interaction Hamiltonians
$H_{pk}^{(l)}(t)$ where $k=2l-1$ or $2l$ respectively.
Here, $\hbar\Omega_{mn,l}=-MI_{mn}\tilde{I}_{dl}/2$.
The incident driving current is assumed as $I_{dl}\left(  x,t\right)  =
\operatorname{Re}[\tilde{I}_{dl}  e^{-i\omega_{dl}(t-x/v)}]$ where
$l=1,2$, or $3$, and $x$ is the position coordinate.
The incident signal currents are assumed as $I_{pk}\left(  x,t\right)  =\operatorname{Re}
[\tilde{I}_{pk}e^{-i\omega_{k}(t-x/v)}]$ with phase velocity $v$. The parameter $\hat{I}$ is the loop current of the flux qubit circuit and $I_{mn}$ is its matrix element in the energy basis.}
\begin{ruledtabular}
\begin{tabular}{c|c |c}
Driving type & Interaction Hamiltonian with the driving field & Interaction Hamiltonian with the probe field \\
\hline
Type (1) & $H_{R}^{(1)}(t)=\hbar\left[\Omega_{21,1}\,\sigma_{21}\exp(-i\omega_{d1}t)+\text{H.c.}\right] $ &
$\begin{array}{c}
H_{p1}^{(1)}(t)=-M\hat{I}I_{p1}(0,t)\\
H_{p2}^{(1)}(t)=-M\hat{I}I_{p2}(0,t)\end{array}$\\
\hline
Type (2) & $H_{R}^{(2)}(t)=\hbar\left[\Omega_{32,2}\,\sigma_{32}\exp(-i\omega_{d2}t)+\text{H.c.}\right]  $&
$\begin{array}{c}
H_{p3}^{(2)}(t)=-M\hat{I}I_{p3}(0,t)\\
H_{p4}^{(2)}(t)=-M\hat{I}I_{p4}(0,t)\end{array}$\\
\hline
Type (3) & $H_{R}^{(3)}(t)=\hbar\left[\Omega_{31,3}\,\sigma_{31}\exp(-i\omega_{d3}t)+\text{H.c.}\right] $ &
$\begin{array}{c}
H_{p5}^{(3)}(t)=-M\hat{I}I_{p5}(0,t)\\
H_{p6}^{(3)}(t)=-M\hat{I}I_{p6}(0,t)\end{array}$
\end{tabular}
\end{ruledtabular}
\end{table*}
\endgroup
\end{widetext}

\section{Microwave attenuation and frequency conversions for driving type
(1)\label{sec:d12}}

\subsection{Hamiltonian reduction}

In the driving type (1), the Hamiltonian $H_{R}^{(1)}(t)$ can be given as
\begin{equation}
H_{R}^{(1)}(t)=\hbar\Omega_{21,1}\,\sigma_{21}\exp(-i\omega_{d1}t)+\text{H.c.}
\label{eq:2}%
\end{equation}
with $\Omega_{21,1}=-MI_{21}\tilde{I}_{d1}/2.$ Here $M$ is the mutual
inductance, and $I_{mn}$ is the matrix element of the loop current $\hat{I}$
of the three-level superconducting flux qubit circuit. The incident driving
current is assumed as $I_{d1}\left(  x,t\right)  =\operatorname{Re}[\tilde
{I}_{d1}e^{-i\omega_{d1}\left(  t-x/v\right)  }]$ with phase velocity $v$. We
assume that the three-level system is placed at the position of $x=0$ and also
$\Omega_{21,1}$ is a real number without loss of generality.

Corresponding to the Hamiltonian in Eq.~(\ref{eq:2}), the Hamiltonian
$H_{pk}^{\left(  1\right)  }(t)$ with $k=1$ or 2 are respectively
\begin{align}
H_{p1}^{(1)}(t)  &  =-M\hat{I}I_{p1}\left(  0,t\right)  ,\label{eq:3}\\
H_{p2}^{(1)}(t)  &  =-M\hat{I}I_{p2}\left(  0,t\right)  , \label{eq:4}%
\end{align}
without the rotating wave approximation (RWA). The Hamiltonian in
Eq.~(\ref{eq:H_Total}) with $H_{R}^{\left(  1\right)  }\left(  t\right)  $ in
Eq.~(\ref{eq:2}) and $H_{p1}^{\left(  1\right)  }\left(  t\right)  $ in
Eq.~(\ref{eq:3}) describes the frequency down conversion as shown in the up
panel of Fig.~\ref{fig1}(b). However, the Hamiltonian $H_{p2}^{(1)}(t)$ can be
written as for that the signal field is applied to the energy levels
$|2\rangle$ and $|3\rangle$. That is, the Hamiltonian in Eq.~(\ref{eq:H_Total}%
) with $H_{R}^{\left(  1\right)  }\left(  t\right)  $ in Eq.~(\ref{eq:2}) and
$H_{p2}^{\left(  1\right)  }\left(  t\right)  $ in Eq.~(\ref{eq:4}) describes
the frequency up conversion as shown in the down panel of Fig.~\ref{fig1}(b).
The incident signal currents are assumed as $I_{pk}\left(  x,t\right)
=\operatorname{Re}[\tilde{I}_{pk}e^{-i\omega_{k}\left(  t-x/v\right)  }]$ with
$k=1$ or $k=2$.

To remove the time dependence of $H^{\left(  1\right)  }$, we now use a
unitary transformation $U_{d}^{\left(  1\right)  }=\exp\left(  -i\omega
_{d1}t\sigma_{22}\right)  $. Then at a frame rotating, we get an effective
Hamiltonian%
\begin{align}
H_{\text{eff}}^{(1)} &  =\hbar\Delta_{21,1}\sigma_{22}+\hbar\omega_{31}%
\sigma_{33}+\hbar\Omega_{21,1}\,\left(  \sigma_{21}+\sigma_{12}\right)
\label{eq:H_eff_l}\\
&  -M\hat{I}^{(1)}\left(  t\right)  I_{pk}\left(  0,t\right)  -M\hat{I}%
^{(1)}\left(  t\right)  (I_{L}\left(  0,t\right)  +I_{R}\left(  0,t\right)
),\nonumber
\end{align}
with driving detuning $\Delta_{21,1}=\omega_{21}-\omega_{d1}$ and the
transformed loop current $\hat{I}^{\left(  1\right)  }\left(  t\right)
=U_{d}^{\left(  1\right)  \dag}\hat{I}U_{d}^{\left(  1\right)  }$. Since we
have assumed a strong driving strength $\Omega_{mn,l}$, it is reasonable to
work in the eigen basis of the first three terms of Eq.~(\ref{eq:H_eff_l}).
For this consideration, we apply to $H_{\text{eff}}^{(1)}$ a unitary
transformation $U_{r}^{\left(  1\right)  }=\exp\left[  -i\theta_{1}\left(
-i\sigma_{21}+i\sigma_{12}\right)  /2\right]  $ with $\tan\theta_{1}%
=2\Omega_{21,1}/\Delta_{21,1}$, yielding%
\begin{equation}
\bar{H}_{\text{eff}}^{(1)}=U_{r}^{\left(  1\right)  \dag}H_{\text{eff}}%
^{(1)}U_{r}^{\left(  1\right)  }=H_{S}^{\left(  1\right)  }+\bar{H}%
_{pk}^{\left(  1\right)  }+H_{T}^{\left(  1\right)  }.\label{eq:Hbareff1}%
\end{equation}
The symbols $H_{S}^{\left(  1\right)  }$, $\bar{H}_{pk}^{\left(  1\right)  }$,
$H_{T}^{\left(  1\right)  }$ respectively take the following forms, i.e.,%
\begin{align}
H_{S}^{\left(  1\right)  } &  =\hbar\omega_{1}^{\left(  1\right)  }\sigma
_{11}+\hbar\omega_{2}^{\left(  1\right)  }\sigma_{22}+\hbar\omega_{3}^{\left(
1\right)  }\sigma_{33},\label{eq:HS1}\\
\bar{H}_{pk}^{\left(  1\right)  } &  =-M\bar{I}^{\left(  1\right)  }\left(
t\right)  I_{pk}\left(  0,t\right)  ,\\
H_{T}^{\left(  1\right)  } &  =-M\bar{I}^{\left(  1\right)  }\left(  t\right)
\left(  I_{L}\left(  0,t\right)  +I_{R}\left(  0,t\right)  \right)  .
\end{align}
Note that $\bar{I}^{\left(  1\right)  }\left(  t\right)  =$ $U_{r}^{\left(
1\right)  \dag}\hat{I}^{\left(  1\right)  }\left(  t\right)  U_{r}^{\left(
1\right)  }$ is another transformed loop current. Its matrix elements $\bar
{I}_{mn}^{\left(  1\right)  }\left(  t\right)  $ have been listed in
Appendix.~\ref{append:d12}. The Hamiltonian $H_{S}^{\left(  1\right)  }$ is
defined as the system Hamiltonian originating from the first three terms of
$H_{\text{eff}}^{(1)}$ in Eq.~(\ref{eq:H_eff_l}). In Eq.~(\ref{eq:HS1}), the
eigen frequencies of the system Hamiltonian are respectively represented as%
\begin{align}
\omega_{1}^{\left(  1\right)  } &  =\frac{1}{2}\left(  \Delta_{21,1}%
-\sqrt{4\Omega_{21,1}^{2}+\Delta_{21,1}^{2}}\right)  ,\\
\omega_{2}^{\left(  1\right)  } &  =\frac{1}{2}\left(  \Delta_{21,1}%
+\sqrt{4\Omega_{21,1}^{2}+\Delta_{21,1}^{2}}\right)  ,\\
\omega_{3}^{\left(  1\right)  } &  =\omega_{31}.
\end{align}
The Hamiltonian $\bar{H}_{pk}^{\left(  1\right)  }$ is a small quantity
compared to $H_{S}^{\left(  1\right)  }$ and hence will be treated as a
perturbation in the following discussions. And $H_{T}^{\left(  1\right)  }$
represents the interaction between the system and 1D open space. With fast
oscillating terms neglected, $\bar{H}_{pk}^{\left(  1\right)  }$ can be
further reduced to%
\begin{align}
\bar{H}_{p1} &  =\hbar\varepsilon_{31,1}e^{-i\omega_{1}t}\sigma_{31}%
+\hbar\varepsilon_{32,1}e^{-i\omega_{1}t}\sigma_{32}+\text{h.c.,}\\
\bar{H}_{p2} &  =\hbar\varepsilon_{31,2}e^{-i\omega_{2+}t}\sigma_{31}%
+\hbar\varepsilon_{32,2}e^{-i\omega_{2+}t}\sigma_{32}+\text{h.c..}%
\end{align}
Here, $\omega_{2+}=\omega_{2}+\omega_{d1}$ is the produced sum frequency, and
\begin{align}
\hbar\varepsilon_{31,1} &  =-\frac{1}{2}M\cos\frac{\theta_{1}}{2}\tilde
{I}_{p1}I_{31},\\
\hbar\varepsilon_{32,1} &  =-\frac{1}{2}M\sin\frac{\theta_{1}}{2}\tilde
{I}_{p1}I_{31},\\
\hbar\varepsilon_{31,2} &  =\frac{1}{2}M\sin\frac{\theta_{1}}{2}\tilde{I}%
_{p2}I_{32},\\
\hbar\varepsilon_{32,2} &  =-\frac{1}{2}M\cos\frac{\theta_{1}}{2}\tilde
{I}_{p2}I_{32},
\end{align}
are the coupling energy parameters.

\subsection{Dynamics of the system and its solutions}

Using the detailed parameters of $\bar{I}^{\left(  1\right)  }\left(
t\right)  $ in Appendix.~\ref{append:d12}, we can derive that the reduced
density matrix $\rho$ of the system is governed by the following master
equation~\cite{QN3}
\begin{equation}
\frac{\partial\rho}{\partial t}=\frac{1}{i\hbar}[H_{S}^{\left(  1\right)
}+\bar{H}_{pk}^{\left(  1\right)  },\rho]+\mathcal{L}\left[  \rho\right]
.\label{eq:6}%
\end{equation}
We must mention we work in the picture defined by unitary transformations
$U_{d}^{\left(  1\right)  }$ and $U_{r}^{\left(  1\right)  }$. The dissipation
of the system is described via the Lindblad term
\begin{align}
\mathcal{L}\left[  \rho\right]   &  =\sum_{m}\left(  \sum_{k\neq
m}\mathcal{\gamma}_{km}^{\left(  1\right)  }\rho_{kk}-\sum_{k\neq
m}\mathcal{\gamma}_{mk}^{\left(  1\right)  }\rho_{mm}\right)  \sigma
_{mm}\nonumber\\
&  -\sum_{m\neq n}\frac{1}{2}\Gamma_{mn}^{\left(  1\right)  }\rho_{mn}%
\sigma_{mn}.\label{eq:Lindblad1}%
\end{align}
Here, $\rho_{mn}\equiv\rho_{mn}\left(  t\right)  $ are matrix elements of the
reduced density operator $\rho\left(  t\right)  $. The relaxation and
dephasing rates can be calculated as $\gamma_{mn}^{\left(  1\right)  }%
=\frac{M^{2}}{\hbar Z_{T}}K_{mn}^{\left(  1\right)  }$ and $\Gamma
_{mn}^{\left(  1\right)  }=\frac{M^{2}}{\hbar Z_{T}}\left(  \sum_{k\neq
m}K_{mk}^{\left(  1\right)  }+\sum_{k\neq n}K_{nk}^{\left(  1\right)
}+K_{\phi mn}^{\left(  1\right)  }\right)  $ from hypotheses (1), (2), and (4)
in Sec.~\ref{sec:model}. The explicit expressions of $K_{mn}^{\left(
1\right)  }$ and $K_{\phi mn}^{\left(  1\right)  }$ are given in
Appendix~\ref{append:d12}.

It is not easy to obtain the exact solutions of the nonlinear equations in
Eq.~(\ref{eq:6}) because the steady state response contains infinite
components of different frequencies in nonlinear systems. Thus, as extensively
used method in nonlinear optics~\cite{NonlinearOptics}, we now seek the
solutions of Eq.~(\ref{eq:6}) in the form of a power series expansion in the
magnitude of $\bar{H}_{pk}^{\left(  1\right)  }$, that is, a solution of the
form
\begin{equation}
\rho(t)=\rho^{(0)}+\rho^{(1)}(t)+\cdots+\rho^{(r)}(t)+\cdots, \label{eq:12}%
\end{equation}
for the reduced density matrix $\rho$ of the three-level system. Here,
$\rho^{\left(  0\right)  }$ is the steady state solution when no signal field
is applied to the system. However the $r$th-order reduced density matrix
$\rho^{\left(  r\right)  }(t)$ is proportional to $r$th order of $\bar{H}%
_{pk}^{\left(  1\right)  }$.

In the first order approximation, we have
\begin{align}
\frac{\partial\rho^{\left(  0\right)  }}{\partial t} &  =\frac{1}{i\hbar
}[H_{S}^{\left(  1\right)  },\rho^{\left(  0\right)  }]+\mathcal{L}\left[
\rho^{\left(  0\right)  }\right]  ,\label{eq:ME1_0}\\
\frac{\partial\rho^{\left(  1\right)  }}{\partial t} &  =\frac{1}{i\hbar
}[H_{S}^{\left(  1\right)  },\rho^{\left(  1\right)  }]+\frac{1}{i\hbar}%
[\bar{H}_{pk}^{\left(  1\right)  },\rho^{\left(  0\right)  }]+\mathcal{L}%
\left[  \rho^{\left(  1\right)  }\right]  .\label{eq:ME1_1}%
\end{align}
The solutions of Eq.~(\ref{eq:ME1_0}) is
\begin{align}
\rho_{11}^{\left(  0\right)  } &  =\frac{1}{1+y_{1}^{2}},\\
\rho_{22}^{\left(  0\right)  } &  =\frac{y_{1}^{2}}{1+y_{1}^{2}},
\end{align}
with $y_{1}=\tan^{2}\left(  \theta_{1}/2\right)  $. And the other terms of
$\rho^{\left(  0\right)  }$ are all zeros. Having obtained $\rho^{\left(
0\right)  }$, we can further solve Eq.~(\ref{eq:ME1_1}). When $\bar{H}%
_{pk}^{\left(  1\right)  }$ takes $\bar{H}_{p1}^{\left(  1\right)  }$, we have
the nonzero matrix elements of $\rho^{\left(  1\right)  }$ as follows,%
\begin{align}
\rho_{31}^{\left(  1\right)  } &  =\rho_{13}^{\left(  1\right)  \ast}%
=\frac{-i\varepsilon_{31,1}e^{-i\omega_{1}t}}{i(\omega_{31}^{\left(  1\right)
}-\omega_{1})+\frac{1}{2}\Gamma_{31}}\rho_{11}^{\left(  0\right)
},\label{eq:rho1p1_1}\\
\rho_{32}^{\left(  1\right)  } &  =\rho_{23}^{\left(  1\right)  \ast}%
=\frac{-i\varepsilon_{32,1}e^{-i\omega_{1}t}}{i(\omega_{32}^{\left(  1\right)
}-\omega_{1})+\frac{1}{2}\Gamma_{32}}\rho_{22}^{\left(  0\right)
}.\label{eq:rho1p1_2}%
\end{align}
When $\bar{H}_{pk}^{\left(  1\right)  }$ takes $\bar{H}_{p2}^{\left(
1\right)  }$, we have the nonzero matrix elements of $\rho^{\left(  1\right)
}$ as follows,%
\begin{align}
\rho_{31}^{\left(  1\right)  } &  =\rho_{13}^{\left(  1\right)  \ast}%
=\frac{-i\varepsilon_{31,2}e^{-i\omega_{2+}t}}{i(\omega_{31}^{\left(
1\right)  }-\omega_{2+})+\frac{1}{2}\Gamma_{31}}\rho_{11}^{\left(  0\right)
},\label{eq:rho1p2_1}\\
\rho_{32}^{\left(  1\right)  } &  =\rho_{23}^{\left(  1\right)  \ast}%
=\frac{-i\varepsilon_{32,2}e^{-i\omega_{2+}t}}{i(\omega_{32}^{\left(
1\right)  }-\omega_{2+})+\frac{1}{2}\Gamma_{32}}\rho_{22}^{\left(  0\right)
}.\label{eq:rho1p2_2}%
\end{align}

\subsection{Scattered current}

The noise current ($I_{L}$ and $I_{R}$) will induce the scattered current of
the classical fields through interaction with the three-level system. Using
the input-output theory extensively discussed in
Refs.~\cite{QN1,QN2,QN3,QN4,QN5}, the scattered current at $x=0$ can be
represented by%
\begin{equation}
I_{s1}\left(  0,t\right)  =-\frac{iM}{2Z_{T}}\sum_{mnk}\delta_{mnk}^{\left(
1\right)  }\bar{I}_{mnk}^{\left(  1\right)  }e^{i\nu_{mnk}^{\left(  1\right)
}t}\rho_{nm},\label{eq:Is1}%
\end{equation}
with $\delta_{mnk}^{\left(  1\right)  }=\omega_{mn}^{\left(  1\right)  }%
+\nu_{mnk}^{\left(  1\right)  }$, and $\omega_{mn}^{\left(  1\right)  }%
=\omega_{m}^{\left(  1\right)  }-\omega_{n}^{\left(  1\right)  }$. Here, we
have assumed that the matrix element of $\bar{I}^{\left(  1\right)  }\left(
t\right)  $ is of the form $\bar{I}_{mn}^{\left(  1\right)  }\left(  t\right)
=\sum_{k}\bar{I}_{mnk}^{\left(  1\right)  }e^{i\nu_{mnk}^{\left(  1\right)
}t}$. The scattered current can also be expanded in the order of $\bar{H}%
_{pk}^{\left(  1\right)  }$, i.e., $I_{s1}=\sum_{r=0}^{\infty}$ $I_{s1}%
^{\left(  r\right)  }$. Here, we only care about the linear response in the
expansion of $I_{s1}$, that is,%
\begin{equation}
I_{s1}^{\left(  1\right)  }\left(  0,t\right)  =-\frac{iM}{2Z_{T}}\sum
_{mnk}\delta_{mnk}^{\left(  1\right)  }\bar{I}_{mnk}^{\left(  1\right)
}e^{i\nu_{mnk}^{\left(  1\right)  }t}\rho_{nm}^{\left(  1\right)  }.
\end{equation}

\subsection{Probe type (1)}

When $H_{pk}^{\left(  1\right)  }$ takes $H_{p1}^{\left(  1\right)  },$ using
Eqs.~(\ref{eq:rho1p1_1})-(\ref{eq:rho1p1_2}), we have the linear response as
\begin{equation}
I_{s1}^{\left(  1\right)  }(0,t)=\operatorname{Re}\{\tilde{I}_{s1}(\omega
_{1})e^{-i\omega_{1}t}\}+\operatorname{Re}\{\tilde{I}_{s1}(\omega
_{1-})e^{-i\omega_{1-}t}\}
\end{equation}
where $\omega_{1-}=\omega_{1}-\omega_{d1}$ is the produced difference
frequency. The amplitudes of both frequency components are respectively
denoted as $\tilde{I}_{s1}(\omega_{1})$ and $\tilde{I}_{s1}(\omega_{1-})$. The
gain of the incident current $I_{p1}$ is defined as $G_{1}=1+\tilde{I}%
_{s1}(\omega_{1})/\tilde{I}_{p1}$, and the explicit expression is
\begin{align}
G_{1}= &  1-\frac{M^{2}}{2\hbar Z_{T}}\frac{\rho_{11}^{\left(  0\right)
}\lambda_{31}\cos^{2}\frac{\theta_{1}}{2}}{i\left(  \omega_{31}^{\left(
1\right)  }-\omega_{1}\right)  +\frac{1}{2}\Gamma_{31}^{\left(  1\right)  }%
}\nonumber\\
&  -\frac{M^{2}}{2\hbar Z_{T}}\frac{\rho_{22}^{\left(  0\right)  }\lambda
_{31}\sin^{2}\frac{\theta_{1}}{2}}{i\left(  \omega_{32}^{\left(  1\right)
}-\omega_{1}\right)  +\frac{1}{2}\Gamma_{32}^{\left(  1\right)  }%
}.\label{eq:G1}%
\end{align}
Meanwhile, the corresponding efficiency of frequency down conversion is
defined as $\eta_{1}=\tilde{I}_{s1}(\omega_{1-})/\tilde{I}_{p1}\sqrt
{\omega_{1}/\omega_{1-}}$ since $\left\vert \eta_{1}\right\vert ^{2}$
represents the photon number of frequency $\omega_{1-}$ produced by each
photon of frequency $\omega_{1}$ per unit time. Then, the explicit expression
of $\eta_{1}$ can be reduced to%
\begin{align}
\eta_{1}= &  \frac{M^{2}}{2\hbar Z_{T}}\frac{\rho_{11}^{\left(  0\right)
}\sqrt{\lambda_{32}\lambda_{31}}\cos\frac{\theta_{1}}{2}\sin\frac{\theta_{1}%
}{2}}{i\left(  \omega_{31}^{\left(  1\right)  }-\omega_{1}\right)  +\frac
{1}{2}\Gamma_{31}^{\left(  1\right)  }}\nonumber\\
&  +\frac{M^{2}}{2\hbar Z_{T}}\frac{\rho_{22}^{\left(  0\right)  }%
\sqrt{\lambda_{32}\lambda_{31}}\cos\frac{\theta_{1}}{2}\sin\frac{\theta_{1}%
}{2}}{i\left(  \omega_{32}^{\left(  1\right)  }-\omega_{1}\right)  +\frac
{1}{2}\Gamma_{32}^{\left(  1\right)  }}.\label{eq:eta1}%
\end{align}

The two resonant points of $G_{1\,}$ and $\eta_{1}$ are respectively at
$\omega_{1}=\omega_{31}^{\left(  1\right)  }$ and $\omega_{1}=\omega
_{32}^{\left(  1\right)  }$. As $\Omega_{21,1}$ is assumed larger than the
decay rates, the two resonant points must be well separated. Therefore, we can
determine from Eq.~(\ref{eq:G1}) that the transmitted signal with frequency
$\omega_{1}$ can only be attenuated. At both points, $\left\vert
G_{1}\right\vert $ ($\left\vert \eta_{1}\right\vert )$ reaches their minimum
(maximum) values respectively. To obtain the optimal attenuation or conversion
efficiency, we can first minimize $\Gamma_{31}^{\left(  1\right)  }$ and
$\Gamma_{32}^{\left(  1\right)  }$ where
\begin{align}
\Gamma_{31}^{\left(  1\right)  } &  =\frac{M^{2}}{\hbar Z_{T}}\left(
\lambda_{31}+\lambda_{32}+\sin^{2}\frac{\theta_{1}}{2}\lambda_{21}\right)  ,\\
\Gamma_{32}^{\left(  1\right)  } &  =\frac{M^{2}}{\hbar Z_{T}}\left(
\lambda_{31}+\lambda_{32}+\cos^{2}\frac{\theta_{1}}{2}\lambda_{21}\right)  .
\end{align}
The dephasing rates $\Gamma_{31}^{\left(  1\right)  }$ and $\Gamma
_{32}^{\left(  1\right)  }$ can be further reduced to
\begin{align}
\Gamma_{31}^{\left(  1\right)  } &  =\frac{M^{2}}{\hbar Z_{T}}\left(
\lambda_{31}+\lambda_{32}\right)  ,\\
\Gamma_{32}^{\left(  1\right)  } &  =\frac{M^{2}}{\hbar Z_{T}}\left(
\lambda_{31}+\lambda_{32}\right)  ,
\end{align}
in the limit that $\lambda_{3}=\lambda_{32}/\lambda_{21}\gg1$ and $\lambda
_{2}=\lambda_{31}/\lambda_{21}\gg1$. We thus assume $\lambda_{3}\gg1$ and
$\lambda_{2}\gg1$ in the following discussions of $\eta_{1}$ and $G_{1}$.

\begin{figure}[ptbh]
\includegraphics[width=0.24\textwidth, clip]{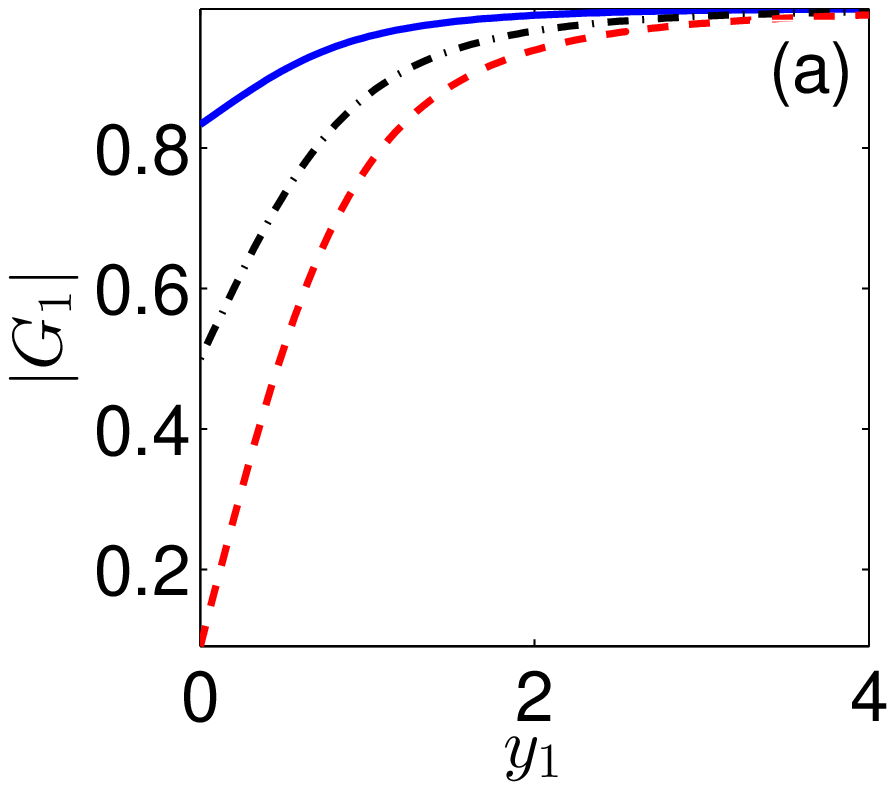}\includegraphics[width=0.24\textwidth, clip]{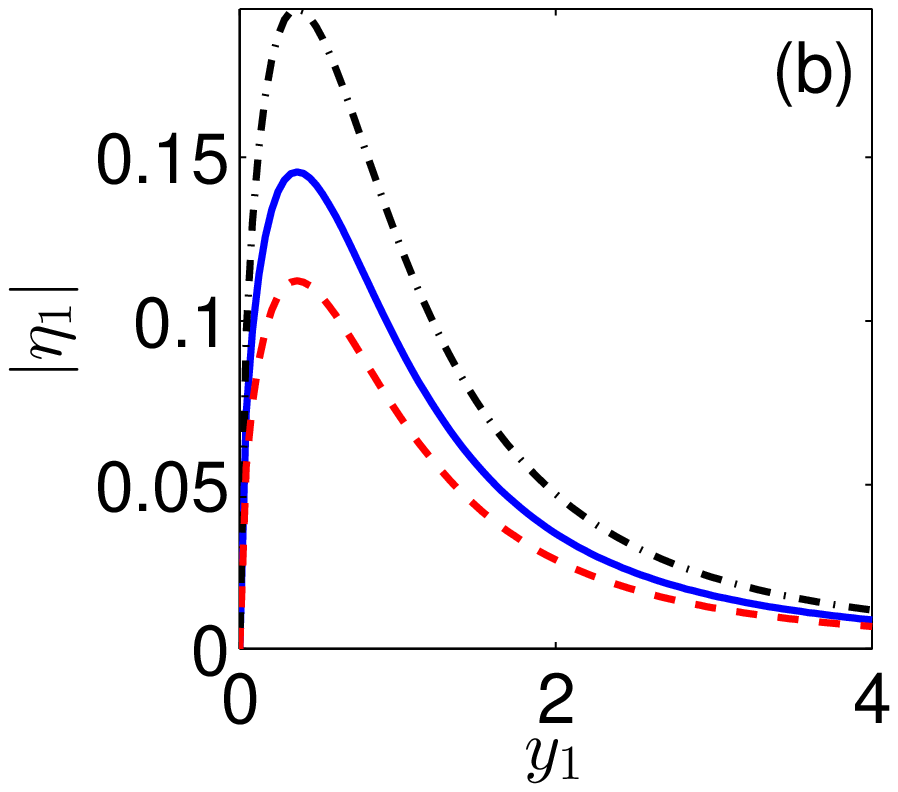}
\caption{(Color online) Probe type (1), driving type (1). The gain (a)
$\left\vert G_{1}\right\vert $ and conversion efficiency (b) $\left\vert
\eta_{1}\right\vert $ plotted as functions of $y_{1}$. Here, we have assumed
that $\omega_{1}=\omega_{31}^{\left(  1\right)  }$ and $\lambda_{2}%
,\lambda_{3}\gg1$. In both (a) and (b), $\lambda_{1}=0.2$ (solid blue), $1$
(dash-dotted black), and $10$ (dashed red), respectively. }%
\label{fig:probe1d1}%
\end{figure}

We now further seek the limitation value of $\left\vert G_{1}\right\vert $
when $\omega_{1}=\omega_{31}^{\left(  1\right)  }$. In this case, $G_{1}$ is
reduced to%
\begin{equation}
G_{1}=1-\frac{\lambda_{1}}{\left(  \lambda_{1}+1\right)  \left(
1+y_{1}\right)  }\frac{1}{1+y_{1}^{2}}, \label{eq:G1_p1}%
\end{equation}
where $\lambda_{1}=\lambda_{31}/\lambda_{32}$. Apparently, when $\lambda
_{1}\gg1$ and $y_{1}\ll1$, the optimal gain for attenuation can reach
$\qquad\qquad$
\begin{equation}
G_{1}=0.
\end{equation}
In the resonant driving case, $y_{1}=1$, and the best gain for attenuation can reach
\begin{equation}
G_{1}=\frac{3}{4},
\end{equation}
under the condition $\lambda_{1}\gg1.$ In Fig.~\ref{fig:probe1d1}(a), $G_{1}$
takes Eq.~(\ref{eq:G1_p1}). We have plotted $\left\vert G_{1}\right\vert $ as
functions of $y_{1}$ when $\lambda_{1}$ takes 0.2, 1, and 10, respectively. It
can be easily seen that $\left\vert G_{1}\right\vert $ will decrease as
$\lambda_{1}$ increases or $y_{1}$ decreases. It is a similar case when
$\omega_{1}=\omega_{32}^{\left(  1\right)  }$, where the gain becomes%
\begin{equation}
G_{1}=1-\frac{\lambda_{1}}{\left(  \lambda_{1}+1\right)  \left(  1+y_{1}%
^{-1}\right)  }\frac{1}{1+y_{1}^{-2}}. \label{eq:G1_p2}%
\end{equation}
The similarity can be easily found between Eqs.~(\ref{eq:G1_p1}) and
(\ref{eq:G1_p2}). Thus, we directly have the optimal gain for attenuation%
\begin{equation}
G_{1}=0,
\end{equation}
when $\lambda_{1}\gg1$ and $y_{1}^{-1}\ll1$. In the resonant driving case,
$y_{1}=1$, and the optimal gain for attenuation can reach
\begin{equation}
G_{1}=\frac{3}{4}%
\end{equation}
with the condition $\lambda_{1}\gg1$. The properties of $\left\vert
G_{1}\right\vert $ when $G_{1}$ takes (\ref{eq:G1_p2}) can also be
investigated through Fig.~\ref{fig:probe1d1}(a).

We now seek the limitation of $|\eta_{1}|$ when $\omega_{1}=\omega
_{31}^{\left(  1\right)  }$. In this case, $\eta_{1}$ is reduced to%
\begin{equation}
\eta_{1}=\frac{\sqrt{\lambda_{1}}}{\left(  \lambda_{1}+1\right)  \left(
1+y_{1}\right)  }\frac{\sqrt{y_{1}}}{1+y_{1}^{2}}. \label{eq:eta1_p1}%
\end{equation}
It can be proved that when $\lambda_{1}=1$ and $y_{1}=0.36349$, the optimal
conversion efficiency reads%
\begin{equation}
\eta_{1}=0.195\,29.
\end{equation}
In the resonant driving case, $y_{1}=1$, and we have the optimal conversion
efficiency
\begin{equation}
\eta_{1}=\frac{1}{8},
\end{equation}
under the condition $\lambda_{1}=1$. In Fig.~\ref{fig:probe1d1}(b), $\eta_{1}$
takes Eq.~(\ref{eq:eta1_p1}\.{)}. We have plotted $\left\vert \eta
_{1}\right\vert $ as functions of $y_{1}$ when $\lambda_{1}$ takes 0.2, 1, and
10, respectively. As $\lambda_{1}$ is given and $y_{1}$ increases, $\left\vert
\eta_{1}\right\vert $ first increases to the maximum point and then fall
towards zero. Given $y_{1}$, the maximum $\left\vert \eta_{1}\right\vert $
emerges at $\lambda_{1}=1$. It is a similar case when $\omega_{1}=\omega
_{32}^{\left(  1\right)  }$, where the conversion efficiency becomes%
\begin{equation}
\eta_{1}=\frac{\sqrt{\lambda_{1}}}{\left(  \lambda_{1}+1\right)  \left(
y_{1}^{-1}+1\right)  }\frac{\sqrt{y_{1}^{-1}}}{\left(  1+y_{1}^{-2}\right)  }.
\label{eq:eta1_p2}%
\end{equation}
The similarity can be easily found between Eqs.~(\ref{eq:eta1_p1}) and
(\ref{eq:eta1_p2}). Thus, the optimal conversion efficiency reads
\begin{equation}
\eta_{1}=0.195\,29,
\end{equation}
when $\lambda_{1}=1$ and $y_{1}^{-1}=0.363\,49$. In the resonant driving case,
i.e., $y_{1}=1$, we have the optimal conversion efficiency
\begin{equation}
\eta_{1}=\frac{1}{8}%
\end{equation}
also under the condition $\lambda_{1}=1$. The properties of $\left\vert
\eta_{1}\right\vert $ when $\eta_{1}$ takes (\ref{eq:eta1_p2}) can also be
investigated through Fig.~\ref{fig:probe1d1}(b).

\subsection{Probe type (2)}

When $H_{pk}^{\left(  1\right)  }$ takes $H_{p2}^{\left(  1\right)  },$ using
Eqs.~(\ref{eq:rho1p2_1})-(\ref{eq:rho1p2_2}), we have the linear response as
\begin{equation}
I_{s1}^{\left(  1\right)  }(0,t)=\operatorname{Re}\{\tilde{I}_{s1}(\omega
_{2})e^{-i\omega_{1}t}\}+\operatorname{Re}\{\tilde{I}_{s1}(\omega
_{2+})e^{-i\omega_{2+}t}\}.
\end{equation}
The amplitudes of both frequency components are respectively $\tilde{I}\left(
\omega_{2}\right)  $ and $\tilde{I}\left(  \omega_{2+}\right)  $. The gain of
the incident current $I_{p2}$ is defined as $G_{2}=1+\tilde{I}_{s1}(\omega
_{2})/\tilde{I}_{p2}$, and the explicit expression is
\begin{align}
G_{2}= &  1-\frac{M^{2}}{2\hbar Z_{T}}\frac{\rho_{11}^{\left(  0\right)
}\lambda_{32}\sin^{2}\frac{\theta_{1}}{2}}{i\left(  \omega_{31}^{\left(
1\right)  }-\omega_{2+}\right)  +\frac{1}{2}\Gamma_{31}^{\left(  1\right)  }%
}\nonumber\\
&  -\frac{M^{2}}{2\hbar Z_{T}}\frac{\rho_{22}^{\left(  0\right)  }\lambda
_{32}\cos^{2}\frac{\theta_{1}}{2}}{i\left(  \omega_{32}^{\left(  1\right)
}-\omega_{2+}\right)  +\frac{1}{2}\Gamma_{32}^{\left(  1\right)  }%
}\label{eq:G2}%
\end{align}
Meanwhile, the corresponding efficiency of frequency down conversion is
defined as $\eta_{2}=\tilde{I}_{s1}\left(  \omega_{2+}\right)  /\tilde{I}%
_{p2}\sqrt{\omega_{2}/\omega_{2+}}$ since $\left\vert \eta_{2}\right\vert
^{2}$ represents the photon number of frequency $\omega_{2+}$ produced by each
photon of frequency $\omega_{2}$ per unit time. The explicit expression of
$\eta_{2}$ is hence%
\begin{align}
\eta_{2}= &  \frac{M^{2}}{2\hbar Z_{T}}\frac{\sqrt{\lambda_{31}\lambda_{32}%
}\rho_{11}^{\left(  0\right)  }\sin\frac{\theta_{1}}{2}\cos\frac{\theta_{1}%
}{2}}{i\left(  \omega_{31}^{\left(  1\right)  }-\omega_{2+}\right)  +\frac
{1}{2}\Gamma_{31}^{\left(  1\right)  }}\nonumber\\
&  -\frac{M^{2}}{2\hbar Z_{T}}\frac{\sqrt{\lambda_{31}\lambda_{32}}\rho
_{22}^{\left(  0\right)  }\sin\frac{\theta_{1}}{2}\cos\frac{\theta_{1}}{2}%
}{i\left(  \omega_{32}^{\left(  1\right)  }-\omega_{2+}\right)  +\frac{1}%
{2}\Gamma_{32}^{\left(  1\right)  }}.\label{eq:eta2}%
\end{align}

The two resonant points of $G_{2\,}$ and $\eta_{2}$ are respectively at
$\omega_{2+}=\omega_{31}^{\left(  1\right)  }$ and $\omega_{2+}=\omega
_{32}^{\left(  1\right)  }$. As we have assumed a sufficiently large
$\Omega_{21,1}$, the two resonant points must be well separated. Therefore, we
can know from Eq.~(\ref{eq:G2}) that the transmitted signal with frequency
$\omega_{2}$ can only be attenuated. At both points, $\left\vert
G_{2}\right\vert $ ($\left\vert \eta_{2}\right\vert )$ reaches their minimum
(maximum) values respectively. As in probe type (1), we should also assume
$\lambda_{3}\gg1$ and $\lambda_{2}\gg1$ in the following discussions for
achieving optimal $\left\vert G_{2}\right\vert $ and $\left\vert \eta
_{2}\right\vert $.

\begin{figure}[ptbh]
\includegraphics[width=0.24\textwidth, clip]{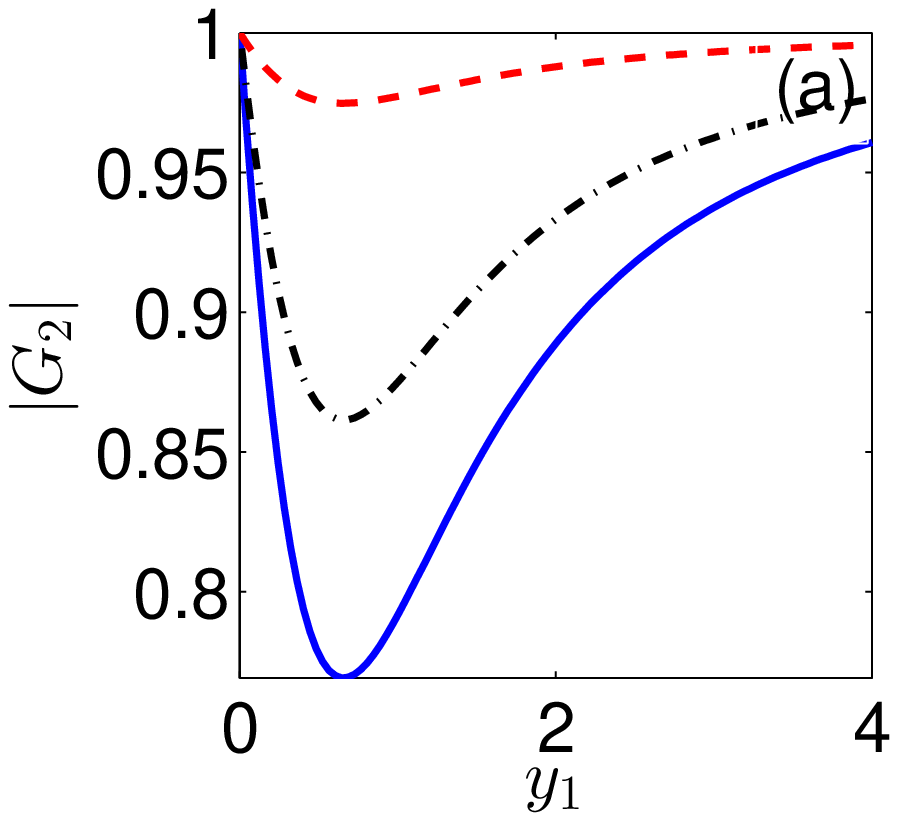}\includegraphics[width=0.24\textwidth, clip]{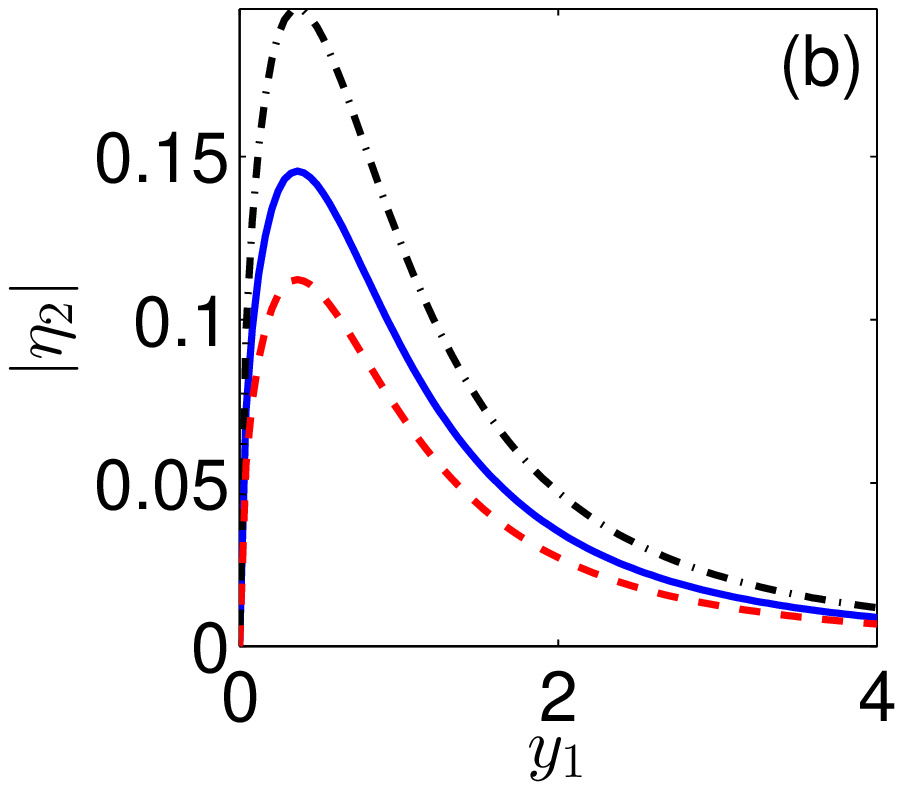}\caption{(Color
online) Probe type (2), driving type (1). The gain (a) $\left\vert
G_{2}\right\vert $ and conversion efficiency (b) $\left\vert \eta
_{2}\right\vert $ plotted as functions of $y_{1}$. Here, we have assumed that
$\omega_{2+}=\omega_{31}^{\left(  1\right)  }$ and $\lambda_{2},\lambda_{3}%
\gg1$. In both (a) and (b), $\lambda_{1}=0.2$ (solid blue), $1$ (dash-dotted
black), and $10$ (dashed red), respectively. }%
\label{fig:probe2d1}%
\end{figure}

We now further seek the limitation value of $\left\vert G_{2}\right\vert $
when $\omega_{2+}=\omega_{31}^{\left(  1\right)  }$. In this case, $G_{2}$ is
reduced to
\begin{equation}
G_{2}=1-\frac{y_{1}}{\left(  \lambda_{1}+1\right)  \left(  1+y_{1}\right)
}\frac{1}{1+y_{1}^{2}}. \label{eq:G2_p1}%
\end{equation}
It can be proved that when $\lambda_{1}\ll1$, and $y_{1}=0.657\,30$,
$\allowbreak$we can achieve the optimal gain for attenuation, that is,
\begin{equation}
G_{2}=0.72305.
\end{equation}
In the resonant driving case, i.e., $y_{1}=1$, the optimal gain for attenuation can
reach%
\begin{equation}
G_{2}=\frac{3}{4},
\end{equation}
under the condition $\lambda_{1}\ll1$. In Fig~\ref{fig:probe2d1}(a), $G_{2}$
takes Eq.~(\ref{eq:G2_p1}). We have plotted $\left\vert G_{2}\right\vert $ as
functions of $y_{1}$ when $\lambda_{1}$ takes 0.2, 1, and 10, respectively.
When $y_{1}$ increases, $\left\vert G_{2}\right\vert $ will first decrease
until meeting the minimum point and then switch to increase.$\ $As
$\lambda_{1}$ increases, $\left\vert G_{2}\right\vert $ will also increase. It
is a similar case when $\omega_{2+}=\omega_{32}^{\left(  1\right)  }$, where
the gain becomes
\begin{equation}
G_{2}=1-\frac{y_{1}^{-1}}{\left(  \lambda_{1}+1\right)  \left(  1+y_{1}%
^{-1}\right)  }\frac{1}{1+y_{1}^{-2}}. \label{eq:G2_p2}%
\end{equation}
The similarity can be easily found between Eqs.~(\ref{eq:G2_p1}) and
(\ref{eq:G2_p2}). Thus we can obtain that when $\lambda_{1}\ll1$ and
$y_{1}^{-1}=0.65730$, the optimal gain for attenuation reads%
\begin{equation}
G_{2}=0.723\,05.
\end{equation}
In the resonant driving case, i.e., $y_{1}=1$, we have the optimal gain for attenuation%
\begin{equation}
G_{2}=\frac{3}{4},
\end{equation}
under the condition $\lambda_{1}\ll1$. When $G_{2}$ takes Eq.~(\ref{eq:G2_p2}%
), $\left\vert G_{2}\right\vert $ can also be investigated through
Fig~\ref{fig:probe2d1}(a).

We now seek the limitation of $|\eta_{2}|$ when $\omega_{2+}=\omega
_{31}^{\left(  1\right)  }$. In this case, $\eta_{2}$ is reduced to%
\begin{equation}
\eta_{2}=\frac{\sqrt{\lambda_{1}}}{\left(  \lambda_{1}+1\right)  \left(
1+y_{1}\right)  }\frac{\sqrt{y_{1}}}{\left(  1+y_{1}^{2}\right)
}.\label{eq:eta2_p1}%
\end{equation}
We find that Eqs.~(\ref{eq:eta2_p1}) and (\ref{eq:eta1_p1}) are exactly of the
same form. We thus directly give that when $\lambda_{1}=1$ and $y_{1}%
=0.36349$, the optimal conversion efficiency reads
\begin{equation}
\eta_{2}=0.19529.
\end{equation}
In the resonant driving case, $y_{1}=1$, and the optimal conversion efficiency
reads%
\begin{equation}
\eta_{2}=\frac{1}{8},
\end{equation}
when $\lambda_{1}=1$. It is a similar case when $\omega_{2+}=\omega
_{32}^{\left(  1\right)  }$, where the conversion efficiency becomes%
\begin{equation}
\eta_{2}=-\frac{\sqrt{\lambda_{1}}}{\left(  \lambda_{1}+1\right)  \left(
1+y_{1}^{-1}\right)  }\frac{\sqrt{y_{1}^{-1}}}{1+y_{1}^{-2}}%
.\label{eq:eta2_p2}%
\end{equation}
The similarity can be easily found between Eqs.~(\ref{eq:eta2_p1}) and
(\ref{eq:eta2_p2}). We hence obtain that when $\lambda_{1}=1$ and $y_{1}%
^{-1}=0.363\,49$, the optimal conversion efficiency reads
\begin{equation}
\eta_{2}=0.195\,29.
\end{equation}
In the resonant driving case, $y_{1}=1$, and the optimal conversion efficiency
reads%
\begin{equation}
\eta_{2}=\frac{1}{8},
\end{equation}
when $\lambda_{1}=1$. For completeness, we also plot Fig.~\ref{fig:probe2d1}%
(b) for $\left\vert \eta_{2}\right\vert $. Whether $\eta_{2}$ takes
Eq.~(\ref{eq:eta2_p1}) or (\ref{eq:eta2_p2}), the behaviours of $\left\vert
\eta_{2}\right\vert $ can be investigated through Fig.~\ref{fig:probe2d1}(b).

\section{Microwave attenuation and frequency conversions for driving type
(2)\label{sec:d23}}

\subsection{Hamiltonian reduction}

In the driving type (2), the Hamiltonian $H_{R}^{(2)}(t)$ can be given as
\begin{equation}
H_{R}^{(2)}(t)=\hbar\Omega_{32,2}\exp\left(  -i\omega_{d2}t\right)
\sigma_{32}+\text{H.c}. \label{eq:Hd2}%
\end{equation}
with $\Omega_{32,2}=-MI_{32}\tilde{I}_{d2}/2.$ The incident driving current is
assumed as $I_{d2}\left(  x,t\right)  =\operatorname{Re}[\tilde{I}%
_{d2}e^{-i\omega_{d2}\left(  t-x/v\right)  }]$ with the phase velocity $v$. We
assume that $\Omega_{32,2}$ is a real number without loss of generality.

Corresponding to the Hamiltonian in Eq.~(\ref{eq:Hd2}), the Hamiltonian
$H_{pk}^{\left(  2\right)  }(t)$ with $k=3$ or 4 are respectively
\begin{align}
H_{p3}^{(1)}(t)  &  =-M\hat{I}I_{p3}\left(  0,t\right)  ,\label{eq:Hp3}\\
H_{p4}^{(1)}(t)  &  =-M\hat{I}I_{p4}\left(  0,t\right)  , \label{eq:Hp4}%
\end{align}
without RWA. The Hamiltonian in Eq.~(\ref{eq:H_Total}) with $H_{R}^{\left(
2\right)  }\left(  t\right)  $ in Eq.~(\ref{eq:Hd2}) and $H_{p3}^{\left(
2\right)  }\left(  t\right)  $ in Eq.~(\ref{eq:Hp3}) describes the frequency
down conversion as shown in the up panel of Fig.~\ref{fig1}(c). However, the
Hamiltonian $H_{p4}^{(2)}(t)$ can be written as for that the signal field is
applied to the energy levels $|1\rangle$ and $|2\rangle$. That is, the
Hamiltonian in Eq.~(\ref{eq:H_Total}) with $H_{R}^{\left(  2\right)  }\left(
t\right)  $ in Eq.~(\ref{eq:Hd2}) and $H_{p4}^{\left(  2\right)  }\left(
t\right)  $ in Eq.~(\ref{eq:Hp4}) describes the frequency up conversion as
shown in the down panel of Fig.~\ref{fig1}(c). The incident signal currents
are assumed as $I_{pk}\left(  x,t\right)  =\operatorname{Re}[\tilde{I}%
_{pk}e^{-i\omega_{k}\left(  t-x/v\right)  }]$ with $k=3$ or $k=4$.

To remove the time dependence of $H^{\left(  2\right)  }$, we now use a
unitary transformation $U_{d}^{\left(  2\right)  }=\exp\left(  -i\omega
_{d2}t\sigma_{33}\right)  $. Then at a frame rotating, we get an effective
Hamiltonian%
\begin{align}
H_{\text{eff}}^{(2)}= &  \hbar\omega_{21}\sigma_{22}+\hbar\left(  \omega
_{21}+\Delta_{32,2}\right)  \sigma_{33}+\hbar\Omega_{32,2}\left(  \sigma
_{32}+\sigma_{23}\right)  \nonumber\\
&  -M\hat{I}^{\left(  2\right)  }\left(  t\right)  \left(  I_{L}\left(
0,t\right)  +I_{R}\left(  0,t\right)  \right)  \nonumber\\
&  -M\hat{I}^{\left(  2\right)  }\left(  t\right)  I_{pk}\left(  0,t\right)
,\label{eq:Heff2}%
\end{align}
with driving detuning $\Delta_{32,2}=\omega_{32}-\omega_{d2}$ and the
transformed loop current $\hat{I}^{\left(  2\right)  }\left(  t\right)
=U_{d}^{\left(  2\right)  \dag}\hat{I}U_{d}^{\left(  2\right)  }$.
Furthermore, we think driving strengths $\Omega_{mn,l}$ are strong enough
compared to the decay rates of the flux qubit circuit. Then, we should work in
the eigen basis of the first three terms of Eq.~(\ref{eq:Heff2}). Thus, we
apply to $H_{\text{eff}}^{(2)}$ a unitary transformation $U_{r}^{\left(
2\right)  }=\exp\left(  -i\theta_{2}\left(  -i\sigma_{32}+i\sigma_{23}\right)
/2\right)  $ with $\tan\theta_{2}=2\Omega_{32,2}/\Delta_{32,2}$, yielding%
\begin{equation}
\bar{H}_{\text{eff}}^{(2)}=H_{S}^{\left(  2\right)  }+\bar{H}_{pk}^{\left(
2\right)  }+H_{T}^{\left(  2\right)  },\label{eq:Hbareff}%
\end{equation}
where
\begin{align}
H_{S}^{\left(  2\right)  } &  =\hbar\omega_{21}^{\left(  2\right)  }%
\sigma_{22}+\hbar\omega_{31}^{\left(  2\right)  }\sigma_{33},\label{eq:HS2}\\
\bar{H}_{pk}^{\left(  2\right)  } &  =-M\bar{I}^{\left(  2\right)  }\left(
t\right)  I_{pk}\left(  0,t\right)  ,\\
H_{T}^{\left(  2\right)  } &  =-M\bar{I}^{\left(  2\right)  }\left(  t\right)
\left(  I_{L}\left(  0,t\right)  +I_{R}\left(  0,t\right)  \right)  .
\end{align}
Note that $\bar{I}^{\left(  2\right)  }\left(  t\right)  =$ $U_{r}^{\left(
2\right)  \dag}\hat{I}^{\left(  2\right)  }\left(  t\right)  U_{r}^{\left(
2\right)  }$ is another transformed loop current and its matrix elements have
been listed in Appendix.~\ref{append:d23}. Here, $H_{S}^{\left(  2\right)  }$
is treated as the system Hamiltonian originating from the first three terms in
Eq.~(\ref{eq:Heff2}). In Eq.~(\ref{eq:HS2}), the eigen frequencies are
respectively%
\begin{align}
\omega_{21}^{\left(  2\right)  } &  =\omega_{21}+\frac{1}{2}\left(
\Delta_{32,2}-\sqrt{4\Omega_{32,2}^{2}+\Delta_{32,2}^{2}}\right)  ,\\
\omega_{31}^{\left(  2\right)  } &  =\omega_{21}+\frac{1}{2}\left(
\Delta_{32,2}+\sqrt{4\Omega_{32,2}^{2}+\Delta_{32,2}^{2}}\right)  .
\end{align}
The Hamiltonian $\bar{H}_{pk}^{\left(  2\right)  }$ is a small quantity
compared to $H_{S}^{\left(  2\right)  }$ and hence will be treated as the
perturbation to the system Hamiltonian. Besides, $H_{T}^{\left(  2\right)  }$
determines the dissipation of the system into the 1D open space. Without fast
oscillating terms neglected, $\bar{H}_{pk}^{\left(  2\right)  }$ can be
further reduced to%
\begin{align}
\bar{H}_{p3} &  =\hbar\varepsilon_{21,3}e^{-i\omega_{3-}t}\sigma_{21}%
+\hbar\varepsilon_{31,3}e^{-i\omega_{3-}t}\sigma_{31}+\text{h.c.,}\\
\bar{H}_{p4} &  =\hbar\varepsilon_{21,4}e^{-i\omega_{4}t}\sigma_{21}%
+\hbar\varepsilon_{31,4}e^{-i\omega_{4}t}\sigma_{31}+\text{h.c.,}%
\end{align}
where $\omega_{3-}=\omega_{3}-\omega_{d2}$ is the produced difference
frequency, and the coupling energy parameters are
\begin{align}
\hbar\varepsilon_{21,3} &  =\frac{1}{2}M\sin\frac{\theta_{2}}{2}\tilde{I}%
_{p3}I_{31},\\
\hbar\varepsilon_{31,3} &  =-\frac{1}{2}M\cos\frac{\theta_{2}}{2}\tilde
{I}_{p3}I_{31},\\
\hbar\varepsilon_{21,4} &  =-\frac{1}{2}M\cos\frac{\theta_{2}}{2}\tilde
{I}_{p4}I_{21},\\
\hbar\varepsilon_{31,4} &  =-\frac{1}{2}M\sin\frac{\theta_{2}}{2}\tilde
{I}_{p4}I_{21}.
\end{align}

\subsection{Dynamics of the system and its solutions}

Using the detailed parameters of $\bar{I}^{\left(  2\right)  }\left(
t\right)  $ in Appendix.~\ref{append:d23}, we can derive that the reduced
density matrix $\chi$ of the system is governed by the following master
equation~\cite{QN3}
\begin{equation}
\frac{\partial\chi}{\partial t}=\frac{1}{i\hbar}[H_{S}^{\left(  2\right)
}+\bar{H}_{pk}^{\left(  2\right)  },\chi]+\mathcal{L}\left[  \chi\right]
.\label{eq:ME2}%
\end{equation}
We must mention we work in the picture defined by unitary transformations
$U_{d}^{\left(  1\right)  }$ and $U_{r}^{\left(  1\right)  }$. The dissipation
of the system is described via the Lindblad term
\begin{align}
\mathcal{L}\left[  \chi\right]   &  =\sum_{m}\left(  \sum_{k\neq
m}\mathcal{\gamma}_{km}^{\left(  2\right)  }\chi_{kk}-\sum_{k\neq
m}\mathcal{\gamma}_{mk}^{\left(  2\right)  }\chi_{mm}\right)  \sigma
_{mm}\nonumber\\
&  -\sum_{m\neq n}\frac{1}{2}\Gamma_{mn}^{\left(  2\right)  }\chi_{mn}%
\sigma_{mn}.\label{eq:Lindblad2}%
\end{align}
Here, $\chi_{mn}\equiv\chi_{mn}\left(  t\right)  $ are matrix elements of the
reduced density operator $\chi\left(  t\right)  $. The relaxation and
dephasing rates can be calculated as $\gamma_{mn}^{\left(  2\right)  }%
=\frac{M^{2}}{\hbar Z_{T}}K_{mn}^{\left(  2\right)  }$ and $\Gamma
_{mn}^{\left(  2\right)  }=\frac{M^{2}}{\hbar Z_{T}}\left(  \sum_{k\neq
m}K_{mk}^{\left(  2\right)  }+\sum_{k\neq n}K_{nk}^{\left(  2\right)
}+K_{\phi mn}^{\left(  2\right)  }\right)  $ from hypotheses (1), (2), and (4)
in Sec.~\ref{sec:model}. The explicit expressions of $K_{mn}^{\left(
2\right)  }$ and $K_{\phi mn}^{\left(  2\right)  }$ are given in
Appendix.~\ref{append:d23}.

Then we seek the solutions of Eq.~(\ref{eq:ME2}) in the form of a power series
expansion in the magnitude of $\bar{H}_{pk}^{\left(  2\right)  }$, that is, a
solution of the form
\begin{equation}
\chi(t)=\chi^{(0)}+\chi^{(1)}(t)+\cdots+\chi^{(r)}(t)+\cdots,
\label{eq:ki_expansion}%
\end{equation}
for the reduced density matrix $\chi$ of the three-level system. Here,
$\chi^{\left(  0\right)  }$ is the steady state solution when no signal field
is applied to the system. However the $r$th-order reduced density matrix
$\chi^{\left(  r\right)  }(t)$ is proportional to $r$th order of $\bar{H}%
_{pk}^{\left(  2\right)  }$.

In the first order approximation, we have
\begin{align}
\frac{\partial\chi^{\left(  0\right)  }}{\partial t} &  =\frac{1}{i\hbar
}[H_{S}^{\left(  2\right)  },\chi^{\left(  0\right)  }]+\mathcal{L[}%
\chi^{\left(  0\right)  }],\label{eq:ME2_0}\\
\frac{\partial\chi^{\left(  1\right)  }}{\partial t} &  =\frac{1}{i\hbar
}[H_{S}^{\left(  2\right)  },\chi^{\left(  1\right)  }]+\frac{1}{i\hbar}%
[\bar{H}_{pk}^{\left(  2\right)  },\chi^{\left(  0\right)  }]+\mathcal{L[}%
\chi^{\left(  1\right)  }].\label{eq:ME2_1}%
\end{align}
The solutions of Eq.~(\ref{eq:ME2_0}) is
\begin{equation}
\chi_{11}^{\left(  0\right)  }=1.
\end{equation}
And the other terms of $\chi^{\left(  0\right)  }$ are all zeros. Having
obtained $\chi^{\left(  0\right)  }$, we can future solve Eq.~(\ref{eq:ME2_1}%
). When $\bar{H}_{pk}^{\left(  2\right)  }$ takes $\bar{H}_{p3}^{\left(
2\right)  }$, we have the nonzero matrix elements of $\chi^{\left(  1\right)
}$ as follows,%
\begin{align}
\chi_{21}^{\left(  1\right)  } &  =\chi_{12}^{\left(  1\right)  \ast}%
=-\frac{i\varepsilon_{21,3}e^{-i\omega_{3-}t}}{i\left(  \omega_{21}^{\left(
2\right)  }-\omega_{3-}\right)  +\frac{1}{2}\Gamma_{21}^{\left(  2\right)  }%
},\label{eq:kip3_1}\\
\chi_{31}^{\left(  1\right)  } &  =\chi_{13}^{\left(  1\right)  \ast}%
=-\frac{i\varepsilon_{31,3}e^{-i\omega_{3-}t}}{i\left(  \omega_{31}^{\left(
2\right)  }-\omega_{3-}\right)  +\frac{1}{2}\Gamma_{31}^{\left(  2\right)  }%
}.\label{eq:kip3_2}%
\end{align}
When $\bar{H}_{pk}^{\left(  2\right)  }$ takes $\bar{H}_{p4}^{\left(
2\right)  }$, we have the nonzero matrix elements of $\chi^{\left(  1\right)
}$ as follows,%
\begin{align}
\chi_{21}^{\left(  1\right)  } &  =\chi_{12}^{\left(  1\right)  \ast}%
=-\frac{i\varepsilon_{21,4}e^{-i\omega_{4}t}}{i\left(  \omega_{21}^{\left(
2\right)  }-\omega_{4}\right)  +\frac{1}{2}\Gamma_{21}^{\left(  2\right)  }%
},\label{eq:ki4_1}\\
\chi_{31}^{\left(  1\right)  } &  =\chi_{13}^{\left(  1\right)  \ast}%
=-\frac{i\varepsilon_{31,4}e^{-i\omega_{4}t}}{i\left(  \omega_{31}^{\left(
2\right)  }-\omega_{4}\right)  +\frac{1}{2}\Gamma_{31}^{\left(  2\right)  }%
}.\label{eq:ki4_2}%
\end{align}

\subsection{Scattered current}

The scattered current at $x=0$, similarly to driving type (1), can be
represented by%
\begin{equation}
I_{s2}\left(  0,t\right)  =-\frac{iM}{2Z_{T}}\sum_{mnk}\delta_{mnk}^{\left(
2\right)  }\bar{I}_{mnk}^{\left(  2\right)  }e^{i\nu_{mnk}^{\left(  2\right)
}t}\chi_{nm},\label{eq:Is2}%
\end{equation}
with $\delta_{mnk}^{\left(  2\right)  }=\omega_{mn}^{\left(  2\right)  }%
+\nu_{mnk}^{\left(  2\right)  }$. Here, we have assumed that the matrix
element of $\bar{I}^{\left(  2\right)  }\left(  t\right)  $ is of the form
$\bar{I}_{mn}^{\left(  2\right)  }\left(  t\right)  =\sum_{k}\bar{I}%
_{mnk}^{\left(  2\right)  }e^{i\nu_{mnk}^{\left(  2\right)  }t}$. We hereby
also expand the scattered current as $I_{s2}=\sum_{r=0}^{\infty}$
$I_{s2}^{\left(  r\right)  }$, where $I_{s2}^{\left(  r\right)  }$ is in the
$r$th order of $\bar{H}_{pk}^{\left(  2\right)  }$. In this paper, we only
care about the linear response of $\bar{H}_{pk}^{\left(  2\right)  }$, that
is,%
\begin{equation}
I_{s2}^{\left(  1\right)  }\left(  0,t\right)  =-\frac{iM}{2Z_{T}}\sum
_{mnk}\delta_{mnk}^{\left(  2\right)  }\bar{I}_{mnk}^{\left(  2\right)
}e^{i\nu_{mnk}^{\left(  2\right)  }t}\chi_{nm}^{\left(  1\right)  }.
\end{equation}

\subsection{Probe type (3)}

When $H_{pk}^{\left(  2\right)  }$ takes $H_{p2}^{\left(  2\right)  },$ using
Eqs.~(\ref{eq:ki4_1})-(\ref{eq:ki4_2}), we have the linear response as
\begin{equation}
I_{s2}^{\left(  1\right)  }(0,t)=\operatorname{Re}\{\tilde{I}_{s2}(\omega
_{3})e^{-i\omega_{3}t}\}+\operatorname{Re}\{\tilde{I}_{s2}(\omega
_{3-})e^{-i\omega_{3-}t}\}
\end{equation}
where $\omega_{3-}=\omega_{3}-\omega_{d2}$ is the produced difference
frequency. The amplitudes of both frequency components are respectively
$\tilde{I}_{s2}(\omega_{3})$ and $\tilde{I}_{s2}(\omega_{3-})$. The gain of
the incident current $I_{p3}$ is defined as $G_{3}=1+\tilde{I}_{s2}(\omega
_{3})/\tilde{I}_{p3}$, and the explicit expression is
\begin{align}
G_{3}= &  1-\frac{M^{2}}{2\hbar Z_{T}}\frac{\lambda_{31}\sin^{2}\frac
{\theta_{2}}{2}}{i\left(  \omega_{21}^{\left(  2\right)  }-\omega_{3-}\right)
+\frac{1}{2}\Gamma_{21}^{\left(  2\right)  }}\nonumber\\
&  -\frac{M^{2}}{2\hbar Z_{T}}\frac{\lambda_{31}\cos^{2}\frac{\theta}{2}%
}{i\left(  \omega_{31}^{\left(  2\right)  }-\omega_{3-}\right)  +\frac{1}%
{2}\Gamma_{31}^{\left(  2\right)  }}.\label{eq:G3}%
\end{align}
Meanwhile, the corresponding efficiency of frequency down conversion is
defined as $\eta_{3}=\tilde{I}_{s2}(\omega_{3-})/\tilde{I}_{p3}\sqrt
{\omega_{3}/\omega_{3-}}$ since $\left\vert \eta_{3}\right\vert ^{2}$
represents the photon number of frequency $\omega_{3-}$ produced by each
photon of frequency $\omega_{3}$ per unit time. The explicit expression of
$\eta_{3}$ is hence%
\begin{align}
\eta_{3} &  =\frac{M^{2}}{2\hbar Z_{T}}\frac{\sqrt{\lambda_{21}\lambda_{31}%
}\sin\frac{\theta_{2}}{2}\cos\frac{\theta_{2}}{2}}{i\left(  \omega
_{21}^{\left(  2\right)  }-\omega_{3-}\right)  +\frac{1}{2}\Gamma
_{21}^{\left(  2\right)  }}\nonumber\\
&  -\frac{M^{2}}{2\hbar Z_{T}}\frac{\sqrt{\lambda_{21}\lambda_{31}}\sin
\frac{\theta_{2}}{2}\cos\frac{\theta_{2}}{2}}{i\left(  \omega_{31}^{\left(
2\right)  }-\omega_{3-}\right)  +\frac{1}{2}\Gamma_{31}^{\left(  2\right)  }%
}.\label{eq:eta3}%
\end{align}

The two resonant points of $G_{3\,}$ and $\eta_{3}$ are respectively at
$\omega_{3-}=\omega_{21}^{\left(  2\right)  }$ and $\omega_{3-}=\omega
_{31}^{\left(  2\right)  }$. As we have assumed a sufficiently large
$\Omega_{32,2}$, the two resonant points must be well separated. Therefore, we
can determine from Eq.~(\ref{eq:G3}) that the transmitted signal with
frequency $\omega_{3}$ can only be attenuated. At both points, $\left\vert
G_{3}\right\vert $ ($\left\vert \eta_{3}\right\vert )$ reaches their minimum
(maximum) values respectively. To obtain the optimal attenuation or conversion
efficiency, we can first minimize $\Gamma_{21}^{\left(  2\right)  }$ and
$\Gamma_{31}^{\left(  2\right)  }$ where
\begin{align}
\Gamma_{21}^{\left(  2\right)  } &  =\frac{M^{2}}{\hbar Z_{T}}\left[
\lambda_{21}\cos^{2}\frac{\theta_{2}}{2}+\left(  \lambda_{31}+\lambda
_{32}\right)  \sin^{2}\frac{\theta_{2}}{2}\right]  ,\\
\Gamma_{31}^{\left(  2\right)  } &  =\frac{M^{2}}{\hbar Z_{T}}\left[
\lambda_{21}\sin^{2}\frac{\theta_{2}}{2}+\left(  \lambda_{31}+\lambda
_{32}\right)  \cos^{2}\frac{\theta_{2}}{2}\right]  .
\end{align}
The dephasing rates $\Gamma_{21}^{\left(  2\right)  }$ and $\Gamma
_{31}^{\left(  2\right)  }$ can be further reduced to
\begin{align}
\Gamma_{21}^{\left(  2\right)  } &  =\frac{M^{2}}{\hbar Z_{T}}\left(
\lambda_{21}\cos^{2}\frac{\theta_{2}}{2}+\lambda_{31}\sin^{2}\frac{\theta_{2}%
}{2}\right)  ,\\
\Gamma_{31}^{\left(  2\right)  } &  =\frac{M^{2}}{\hbar Z_{T}}\left(
\lambda_{21}\sin^{2}\frac{\theta_{2}}{2}+\lambda_{31}\cos^{2}\frac{\theta_{2}%
}{2}\right)  .
\end{align}
in the limit that $\lambda_{1}=\lambda_{31}/\lambda_{32}\gg1$ and $\lambda
_{3}=\lambda_{32}/\lambda_{21}\ll1$. We thus assume $\lambda_{1}\gg1$ and
$\lambda_{3}\ll1$ in the following discussions of $\eta_{3}$ and $G_{3}$.

\begin{figure}[ptbh]
\includegraphics[width=0.24\textwidth, clip]{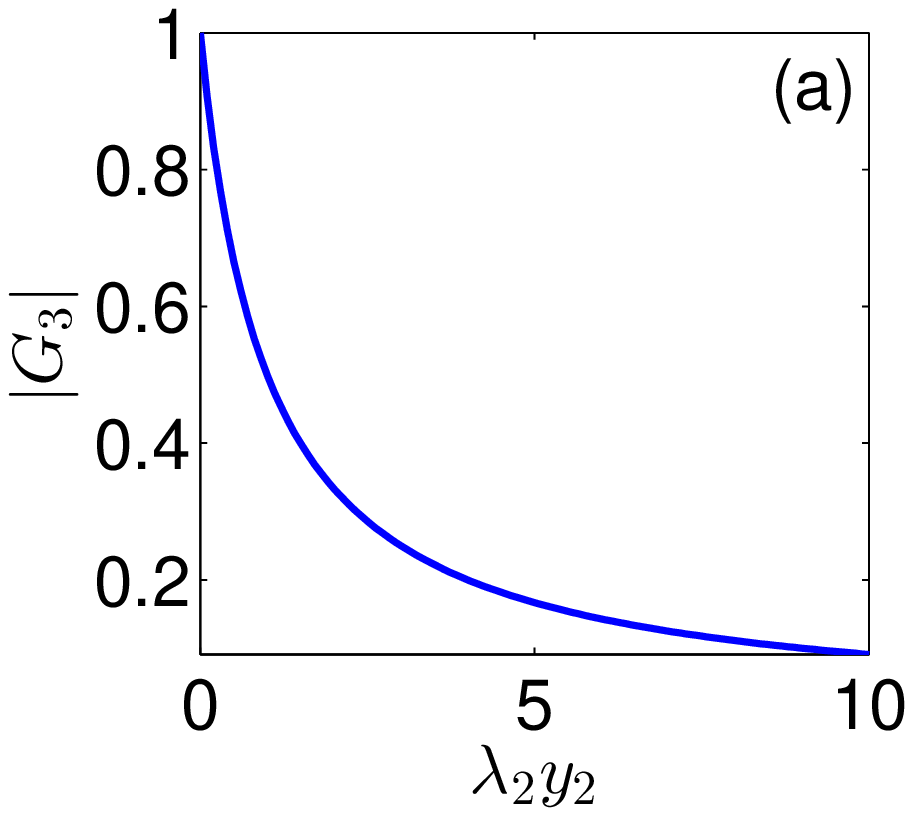}\includegraphics[width=0.24\textwidth, clip]{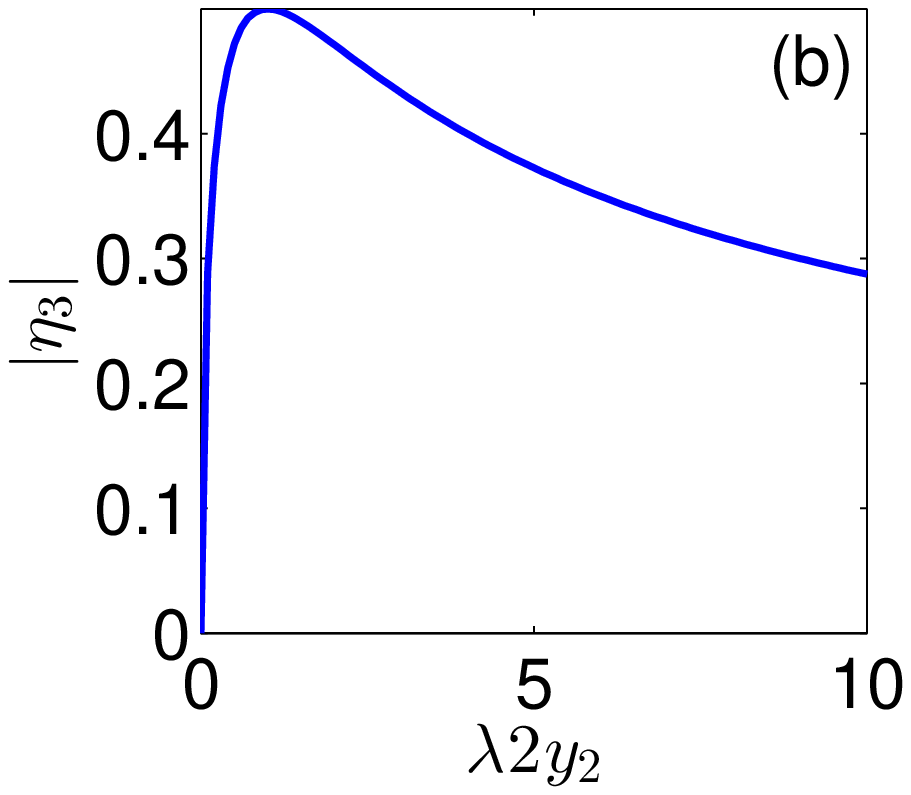}\caption{(Color
online) Probe type (3), driving type (2). The gain (a) $\left\vert
G_{3}\right\vert $ and conversion efficiency (b) $\left\vert \eta
_{3}\right\vert $ plotted as functions of $\lambda_{2}y_{2}$. Here, we have
assumed that $\omega_{3-}=\omega_{21}^{\left(  2\right)  }$, $\lambda_{1}\gg
1$, and $\lambda_{3}\ll1$. }%
\label{fig:probe3}%
\end{figure}

We now further seek the limitation value of $\left\vert G_{3}\right\vert $
when $\omega_{3-}=\omega_{21}^{\left(  2\right)  }$. In this case, $G_{3}$ is
reduced to%
\begin{equation}
G_{3}=1-\frac{\lambda_{2}y_{2}}{1+\lambda_{2}y_{2}},\label{eq:G3_p1}%
\end{equation}
with $y_{2}=\tan^{2}\left(  \theta_{2}/2\right)  $. Apparently, when
$\lambda_{2}y_{2}\gg1$, the optimal gain for attenuation reads$\qquad\qquad$
\begin{equation}
G_{3}=0.
\end{equation}
In the resonant driving case, $y_{2}=1$, and the optimal attenuation $G_{3}=0$
is also achievable with $\lambda_{2}\gg1$. In Fig.~\ref{fig:probe3}(a),
$G_{3}$ takes Eq.~(\ref{eq:G3_p1}). We have plotted $\left\vert G_{3}%
\right\vert $ as a function of $\lambda_{2}y_{2}$. When $\lambda_{2}y_{2}$
increases, $\left\vert G_{3}\right\vert $ also shows decrease towards zero. It
is a similar case when $\omega_{3-}=\omega_{31}^{\left(  2\right)  }$, where
the gain becomes
\begin{equation}
G_{3}=1-\frac{\lambda_{2}y_{2}^{-1}}{1+\lambda_{2}y_{2}^{-1}}.\label{eq:G3_p2}%
\end{equation}
The similarity can be easily found between Eqs.~(\ref{eq:G3_p1}) and
(\ref{eq:G3_p2}). Thus, the optimal gain for attenuation reads%
\begin{equation}
G_{3}=0.
\end{equation}
when the condition $\lambda_{2}y_{2}^{-1}\gg1$ is satisfied. In the resonant
driving case, $y_{2}=1$, and the optimal gain for attenuation $G_{3}=0$ is also
achieved with $\lambda_{2}\gg1$. When $G_{3}$ takes Eq.~(\ref{eq:G3_p2}), the
properties of $\left\vert G_{3}\right\vert $ can also be investigated through
Fig.~\ref{fig:probe3}(a).

We now seek the limitation of $|\eta_{3}|$ when $\omega_{3-}=\omega
_{21}^{\left(  2\right)  }$. In this case, $\eta_{3}$ is reduced to%
\begin{equation}
\eta_{3}=\frac{\sqrt{\lambda_{2}y_{2}}}{1+\lambda_{2}y_{2}}.\label{eq:eta3_p1}%
\end{equation}
When $\lambda_{2}y_{2}=1$, the optimal conversion efficiency reads%
\begin{equation}
\eta_{3}=\frac{1}{2}.
\end{equation}
In the resonant driving case, $y_{2}=1$, and the optimal conversion efficiency
$\eta_{3}=1/2$ is also achievable when $\lambda_{2}=1$. In
Fig.~\ref{fig:probe3}(b), $\eta_{3}$ takes Eq.~(\ref{eq:eta3_p1}). We have
plotted $\left\vert \eta_{3}\right\vert $ as a function of $\lambda_{2}y_{2}$.
When $\lambda_{2}y_{2}$ increases, we find that $\left\vert \eta
_{3}\right\vert $ first increases to the optimal point and then falls towards
zero. It is a similar case when $\omega_{3-}=\omega_{31}^{\left(  2\right)  }%
$, where the conversion efficiency reads
\begin{equation}
\eta_{3}=-\frac{\sqrt{\lambda_{2}y_{2}^{-1}}}{1+\lambda_{2}y_{2}^{-1}%
}.\label{eq:eta_p2}%
\end{equation}
The similarity can be easily found between Eqs.~(\ref{eq:eta_p2}) and
(\ref{eq:eta3_p1}). When $\lambda_{2}y_{2}^{-1}=1$, we have the optimal
conversion efficiency
\begin{equation}
\eta_{3}=-\frac{1}{2}.
\end{equation}
In the resonant driving case, $y_{2}=1$, and the optimal conversion efficiency
$\eta_{3}=-1/2$ is also accessible with $\lambda_{2}=1$. When $\eta_{3}$ takes
Eq.~(\ref{eq:eta_p2}), the property of $\left\vert \eta_{3}\right\vert $ can
be similarly investigated through Fig.~\ref{fig:probe3}%
(b).

\subsection{Probe type (4)}

When $H_{pk}^{\left(  2\right)  }$ takes $H_{p4}^{\left(  2\right)  },$ using
Eqs.~(\ref{eq:ki4_1})-(\ref{eq:ki4_2}), we have the linear response as
\begin{equation}
I_{s2}^{\left(  1\right)  }(0,t)=\operatorname{Re}\{\tilde{I}_{s2}(\omega
_{4})e^{-i\omega_{4}t}\}+\operatorname{Re}\{\tilde{I}_{s2}(\omega
_{4+})e^{-i\omega_{4+}t}\}.
\end{equation}
The amplitudes of both frequency components are respectively $\tilde{I}\left(
\omega_{4}\right)  $ and $\tilde{I}\left(  \omega_{4+}\right)  $. The gain of
the incident current $I_{p4}$ is defined as $G_{4}=1+\tilde{I}_{s2}\left(
\omega_{4}\right)  /\tilde{I}_{p4}$, and the explicit expression is
\begin{align}
G_{4}= &  1-\frac{M^{2}}{2\hbar Z_{T}}\frac{\lambda_{21}\cos^{2}\frac
{\theta_{2}}{2}}{i\left(  \omega_{21}^{\left(  2\right)  }-\omega_{4}\right)
+\frac{1}{2}\Gamma_{21}^{\left(  2\right)  }}\nonumber\\
&  -\frac{M^{2}}{2\hbar Z_{T}}\frac{\lambda_{21}\sin^{2}\frac{\theta_{2}}{2}%
}{i\left(  \omega_{31}^{\left(  2\right)  }-\omega_{4}\right)  +\frac{1}%
{2}\Gamma_{31}^{\left(  2\right)  }}\label{eq:G4}%
\end{align}
Meanwhile, the corresponding efficiency of frequency down conversion is
defined as $\eta_{4}=\tilde{I}\left(  \omega_{4+}\right)  /\tilde{I}_{p4}%
\sqrt{\omega_{4}/\omega_{4+}}$ since $\left\vert \eta_{4}\right\vert ^{2}$
represents the photon number of frequency $\omega_{4+}$ produced by each
photon of frequency $\omega_{4}$ per unit time. The explicit expression of
$\eta_{4}$ is hence%
\begin{align}
\eta_{4}= &  \frac{M^{2}}{2\hbar Z_{T}}\frac{\sqrt{\lambda_{21}\lambda_{31}%
}\sin\frac{\theta_{2}}{2}\cos\frac{\theta_{2}}{2}}{i\left(  \omega
_{21}^{\left(  2\right)  }-\omega_{4}\right)  +\frac{1}{2}\Gamma_{21}^{\left(
2\right)  }}\nonumber\\
&  -\frac{M^{2}}{2\hbar Z_{T}}\frac{\sqrt{\lambda_{21}\lambda_{31}}\sin
\frac{\theta_{2}}{2}\cos\frac{\theta_{2}}{2}}{i\left(  \omega_{31}^{\left(
2\right)  }-\omega_{4}\right)  +\frac{1}{2}\Gamma_{31}^{\left(  2\right)  }%
}.\label{eq:eta4}%
\end{align}

The two resonant points of $G_{4\,}$ and $\eta_{4}$ are respectively at
$\omega_{4+}=\omega_{21}^{\left(  2\right)  }$ and $\omega_{4+}=\omega
_{31}^{\left(  2\right)  }$. As we have assumed a sufficiently large
$\Omega_{32,2}$, the two points must be well separated. Therefore, we can
determine from Eq.~(\ref{eq:G4}) that the transmitted signal with frequency
$\omega_{4}$ can only be attenuated. At both points, $\left\vert
G_{4}\right\vert $ ($\left\vert \eta_{4}\right\vert $) reaches their minimum
(maximum) values respectively. In the following discussions, we will similarly
assume $\lambda_{1}\gg1$ and $\lambda_{3}\ll1$ just as in probe type (3).

\begin{figure}[ptbh]
\includegraphics[width=0.24\textwidth, clip]{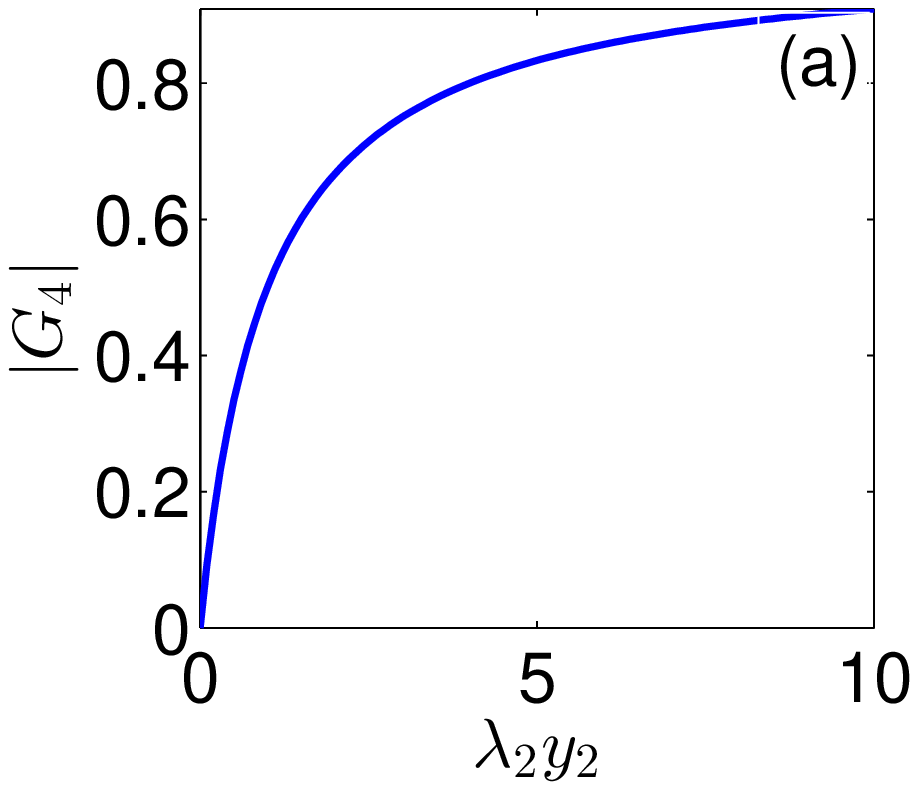}\includegraphics[width=0.24\textwidth, clip]{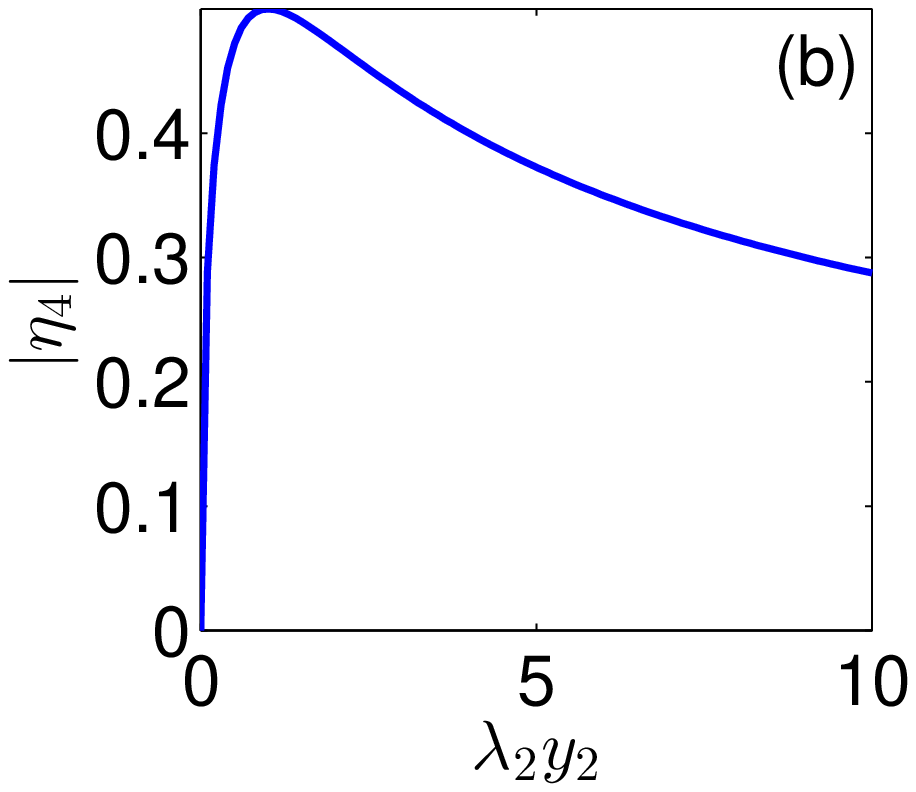}\caption{(Color
online) Probe type (4), driving type (2). The gain (a) $\left\vert
G_{4}\right\vert $ and conversion efficiency (b) $\left\vert \eta
_{4}\right\vert $ plotted as functions of $\lambda_{2}y_{2}$. Here, we have
assumed that $\omega_{4}=\omega_{21}^{\left(  2\right)  }$, $\lambda_{1}\gg1$,
and $\lambda_{3}\ll1$. }%
\label{fig:probe4}%
\end{figure}

We now further seek the limitation value of $\left\vert G_{4}\right\vert $
when $\omega_{4+}=\omega_{21}^{\left(  2\right)  }$. In this case, $G_{4}$ is
hence reduced to%
\begin{equation}
G_{4}=1-\frac{1}{1+\lambda_{2}y_{2}}.\label{eq:G4_p1}%
\end{equation}
When $\lambda_{2}y_{2}\ll1$, we have the optimal gain for attenuation reading
\begin{equation}
G_{4}=0.
\end{equation}
In the resonant driving case, $y_{2}=1$, and the optimal gain for attenuation $G_{4}=0$
is also achievable when $\lambda_{2}\ll1$. In Fig.~\ref{fig:probe4}(a),
$G_{4}$ takes \ref{eq:G4}. We have plotted $\left\vert G_{4}\right\vert $ as
the function of $\lambda_{2}y_{2}$. When $\lambda_{2}y_{2}$ increase,
$\left\vert G_{4}\right\vert $ exhibits increase towards one. It is a similar
case when $\omega_{4+}=\omega_{31}^{\left(  2\right)  }$, where the gain
becomes%
\begin{equation}
G_{4}=1-\frac{1}{1+\lambda_{2}y_{2}^{-1}}.\label{eq:G4_p2}%
\end{equation}
The similarity can be easily seen between Eqs.~(\ref{eq:G4_p2}) and
(\ref{eq:G4_p1}). Thus when $\lambda_{2}y_{2}^{-1}\ll1$, the optimal gain
attenuation reads
\begin{equation}
G_{4}=0.
\end{equation}
In the resonant driving case, $y_{2}=1$, and the optimal attenuation is also
$G_{4}=0$ with $\lambda_{2}\ll1$. When $G_{4}$ takes Eq.~(\ref{eq:G4_p2}), the
behaviours of $\left\vert G_{4}\right\vert $ can be similarly explained
through Fig.~\ref{fig:probe4}(a).

We now seek the limitation of $|\eta_{4}|$ when $\omega_{4+}=\omega
_{21}^{\left(  2\right)  }$. In this case, the conversion efficiency $\eta
_{4}$ is reduced to%
\begin{equation}
\eta_{4}=\frac{\sqrt{\lambda_{2}y_{2}}}{1+\lambda_{2}y_{2}}.\label{eq:eta4_p1}%
\end{equation}
Apparently, Eqs.~(\ref{eq:eta4_p1}) and (\ref{eq:eta3_p1}) are of the same
form. We hence directly have that when $\lambda_{2}y_{2}=1,$ the optimal
conversion efficiency reads%
\begin{equation}
\eta_{4}=\frac{1}{2}.
\end{equation}
In the resonant driving case, $y_{2}=1$, and the optimal conversion efficiency
is also $\eta_{4}=1/2$ with $\lambda_{2}=1$. It is a similar case when
$\omega_{4+}=\omega_{31}^{\left(  1\right)  }$, where%
\begin{equation}
\eta_{4}=-\frac{\sqrt{\lambda_{2}y_{2}^{-1}}}{1+\lambda_{2}y_{2}^{-1}%
}.\label{eq:eta4_p2}%
\end{equation}
When $\lambda_{2}y_{2}^{-1}=1$, the optimal conversion efficiency reads%
\begin{equation}
\eta_{4}=-\frac{1}{2}.
\end{equation}
In the resonant driving case, $y_{2}=1$, and the optimal conversion efficiency
is also $\eta_{4}=-1/2$ with $\lambda_{2}=1$. For completeness, we also plot
Fig.~\ref{fig:probe4}(b) for $\left\vert \eta_{4}\right\vert $. Whether
$\eta_{4}$ takes Eq.~(\ref{eq:eta4_p1}) or (\ref{eq:eta4_p2}), the behaviours
of $\left\vert \eta_{4}\right\vert $ can be investigated through
Fig.~\ref{fig:probe4}(b). 

\section{Microwave amplification, attenuation, and frequency conversions in
the driving type (3)\label{sec:d13}}

According to the analysis on the frequency conversion and the properties of
the output signal field in both driving types (1) and (2). We find that the
signal fields can only be attenuated and not be amplified in both cases.
However the frequency conversion can be realized. We now study the frequency
conversion and the amplification (or attenuation) of incident signal field for
the driving type (3), in which the energy levels $|1\rangle$ and $|3\rangle$
are driven by the strong pump field, and the signal field is used to couple
either the energy levels $|1\rangle$ and $|2\rangle$ or the energy levels
$|2\rangle$ and $|3\rangle$. In this driving case, we find that the incident
signal field can be amplified. The detailed analysis is given below.

\subsection{Hamiltonian reduction}

In the driving type (3), the Hamiltonian $H_{R}^{(3)}(t)$ can be given as
\begin{equation}
H_{R}^{(3)}(t)=\hbar\Omega_{31,3}\exp\left(  -i\omega_{d3}t\right)
\sigma_{31}+\text{H.c}. \label{eq:Hd3}%
\end{equation}
with $\Omega_{31,3}=-MI_{31}\tilde{I}_{d3}/2.$ The incident driving current is
assumed as $I_{d3}\left(  x,t\right)  =\operatorname{Re}[\tilde{I}%
_{d3}e^{-i\omega_{d3}\left(  t-x/v\right)  }]$ with the phase velocity $v$.
The strength $\Omega_{31,3}$ is assumed as a real number without loss of generality.

Corresponding to the Hamiltonian in Eq.~(\ref{eq:Hd3}), the Hamiltonian
$H_{pk}^{\left(  3\right)  }(t)$ with $k=5$ or 6 are respectively
\begin{align}
H_{p5}^{(1)}(t) &  =-M\hat{I}I_{p5}\left(  0,t\right)  ,\label{eq:Hp5}\\
H_{p6}^{(1)}(t) &  =-M\hat{I}I_{p6}\left(  0,t\right)  ,\label{eq:Hp6}%
\end{align}
without RWA. The Hamiltonian in Eq.~(\ref{eq:H_Total}) with $H_{R}%
^{(3)}\left(  t\right)  $ in Eq.~(\ref{eq:Hd3}) and $H_{p5}^{\left(  3\right)
}\left(  t\right)  $ in Eq.~(\ref{eq:Hp5}) describes the frequency up
conversion as shown in the up panel of Fig.~\ref{fig1}(d). However, the
Hamiltonian $H_{p6}^{(3)}(t)$ can be written as for that the signal field is
applied to the energy levels $|2\rangle$ and $|3\rangle$. That is, the
Hamiltonian in Eq.~(\ref{eq:H_Total}) with $H_{R}^{\left(  3\right)  }\left(
t\right)  $ in Eq.~(\ref{eq:Hd2}) and $H_{p4}^{\left(  2\right)  }\left(
t\right)  $ in Eq.~(\ref{eq:Hp6}) describes the frequency down conversion as
shown in the down panel of Fig.~\ref{fig1}(c). The incident signal currents
are assumed as $I_{pk}\left(  x,t\right)  =\operatorname{Re}[\tilde{I}%
_{pk}e^{-i\omega_{k}\left(  t-x/v\right)  }]$ with $k=5$ or $k=6$.

To remove the time dependence of $H^{\left(  3\right)  }$, we now use a
unitary transformation $U_{d}^{\left(  3\right)  }=\exp\left(  -i\omega
_{d3}t\sigma_{33}\right)  $. Then at a frame rotating, we get an effective
Hamiltonian%
\begin{align}
H_{\text{eff}}^{(3)}= &  \hbar\omega_{21}\sigma_{22}+\hbar\Delta_{31,3}%
\sigma_{33}+\hbar\Omega_{31,3}\left(  \sigma_{31}+\sigma_{13}\right)
\nonumber\\
&  -M\hat{I}^{\left(  3\right)  }\left(  t\right)  \left(  I_{L}\left(
0,t\right)  +I_{R}\left(  0,t\right)  \right)  \nonumber\\
&  -M\hat{I}^{\left(  3\right)  }\left(  t\right)  I_{pk}\left(  0,t\right)
,\label{eq:Heff3}%
\end{align}
with driving detuning $\Delta_{31,3}=\omega_{31}-\omega_{d3}$ and loop current
$\hat{I}^{\left(  3\right)  }\left(  t\right)  =U_{d}^{\left(  3\right)  \dag
}\hat{I}U_{d}^{\left(  3\right)  }$. Furthermore, we think driving strengths
$\Omega_{mn,l}$ are strong enough compared to the decay rates of the flux
qubit circuit. Then, we should work in the eigen basis of the first three
terms of Eq.~(\ref{eq:Heff3}). Thus, we apply to $H_{\text{eff}}^{(3)}$ a
unitary transformation $U_{r}^{\left(  3\right)  }=\exp\left(  -i\theta
_{3}\left(  -i\sigma_{31}+i\sigma_{13}\right)  /2\right)  $ with $\tan
\theta_{3}=2\Omega_{31,3}/\Delta_{31,3},$ yielding%
\begin{equation}
\bar{H}_{\text{eff}}^{(3)}=H_{S}^{\left(  3\right)  }+\bar{H}_{pk}^{\left(
3\right)  }+H_{T}^{\left(  3\right)  },
\end{equation}
where
\begin{align}
H_{S}^{\left(  3\right)  } &  =\hbar\omega_{1}^{\left(  3\right)  }\sigma
_{11}+\hbar\omega_{2}^{\left(  3\right)  }\sigma_{22}+\hbar\omega_{3}^{\left(
3\right)  }\sigma_{33},\\
\bar{H}_{pk}^{\left(  3\right)  } &  =-M\bar{I}^{\left(  3\right)  }\left(
t\right)  I_{pk}\left(  0,t\right)  ,\\
H_{T}^{\left(  3\right)  } &  =-M\bar{I}^{\left(  3\right)  }\left(  t\right)
\left(  I_{L}\left(  0,t\right)  +I_{R}\left(  0,t\right)  \right)  .
\end{align}
Here, the loop current $\bar{I}^{\left(  3\right)  }\left(  t\right)  =$
$U_{r}^{\left(  3\right)  \dag}\hat{I}^{\left(  3\right)  }\left(  t\right)
U_{r}^{\left(  3\right)  }$ and its matrix elements have been listed in
Appendix.~\ref{append:d13}. Here, $H_{S}^{\left(  3\right)  }$ is treated as
the system Hamiltonian originating from the first three terms in
Eq.~(\ref{eq:Heff2}). In Eq.~(\ref{eq:HS2}), the eigen frequencies are
respectively%
\begin{align}
\omega_{1}^{\left(  3\right)  } &  =\frac{1}{2}\left(  \Delta_{31,3}%
-\sqrt{4\Omega_{31,1}^{2}+\Delta_{31,3}^{2}}\right)  ,\\
\omega_{2}^{\left(  3\right)  } &  =\omega_{21},\\
\omega_{3}^{\left(  3\right)  } &  =\frac{1}{2}\left(  \Delta_{31,3}%
+\sqrt{4\Omega_{31,1}^{2}+\Delta_{31,3}^{2}}\right)  .
\end{align}
The Hamiltonian $\bar{H}_{pk}^{\left(  3\right)  }$ is a small quantity
compared to $H_{S}^{\left(  3\right)  }$ and hence will be treated as the
perturbation to the system Hamiltonian. Besides, $H_{T}^{\left(  3\right)  }$
determines the dissipation of the system into the 1D open space. With fast
oscillating terms neglected, $\bar{H}_{pk}^{\left(  3\right)  }$ can be
further reduced to%
\begin{align}
\bar{H}_{p5} &  =\hbar\varepsilon_{21,5}e^{-i\omega_{5}t}\sigma_{21}%
+\hbar\varepsilon_{32,5}e^{-i\omega_{5}t}\sigma_{23}+\text{h.c.}\\
\bar{H}_{p6} &  =\hbar\varepsilon_{21,6}e^{i\omega_{6-}t}\sigma_{12}%
+\hbar\varepsilon_{32,6}e^{i\omega_{6-}t}\sigma_{32}+\text{h.c.}%
\end{align}
where $\omega_{6-}=\omega_{d3}-\omega_{6}$ is the produced difference
frequency, and the coupling energy parameters are%
\begin{align}
\hbar\varepsilon_{21,5} &  =\frac{-M\cos\frac{\theta_{3}}{2}\tilde{I}%
_{p5}\left(  0\right)  I_{21}}{2},\\
\hbar\varepsilon_{32,5} &  =\frac{-M\sin\frac{\theta_{3}}{2}\tilde{I}%
_{p5}I_{21}}{2},\\
\hbar\varepsilon_{21,6} &  =\frac{M\sin\frac{\theta_{3}}{2}\tilde{I}%
_{p6}I_{32}}{2},\\
\hbar\varepsilon_{32,6} &  =\frac{-M\cos\frac{\theta_{3}}{2}\tilde{I}%
_{p6}I_{32}}{2}.
\end{align}

\subsection{Dynamics of the system and its solutions}

Using the detailed parameters of $\bar{I}^{\left(  3\right)  }\left(
t\right)  $ in Appendix.~\ref{append:d13}, we can derive that the reduced
density matrix $s$ of the system is governed by the following master
equation~\cite{QN3}
\begin{equation}
\frac{\partial s}{\partial t}=\frac{1}{i\hbar}[H_{S}^{\left(  3\right)  }%
+\bar{H}_{pk}^{\left(  3\right)  },s]+\mathcal{L}\left[  s\right]  .
\end{equation}
We must mention we work in the picture defined by unitary transformations
$U_{d}^{\left(  1\right)  }$ and $U_{r}^{\left(  1\right)  }$. The dissipation
of the system is described via the Lindblad term
\begin{align}
\mathcal{L}\left[  s\right]  = &  \sum_{m}\left(  \sum_{k\neq m}%
\mathcal{\gamma}_{km}^{\left(  3\right)  }s_{kk}-\sum_{k\neq m}\mathcal{\gamma
}_{mk}^{\left(  3\right)  }s_{mm}\right)  \sigma_{mm}\nonumber\\
&  -\sum_{m\neq n}\frac{1}{2}\Gamma_{mn}^{\left(  2\right)  }s_{mn}\sigma
_{mn}.
\end{align}
Here, $s_{mn}\equiv s_{mn}\left(  t\right)  $ are matrix elements of the
reduced density operator $s\left(  t\right)  $. The relaxation and dephasing
rates can be calculated as $\gamma_{mn}^{\left(  3\right)  }=\frac{M^{2}%
}{\hbar Z_{T}}K_{mn}^{\left(  3\right)  }$ and $\Gamma_{mn}^{\left(  3\right)
}=\frac{M^{2}}{\hbar Z_{T}}\left(  \sum_{k\neq m}K_{mk}^{\left(  3\right)
}+\sum_{k\neq n}K_{nk}^{\left(  3\right)  }+K_{\phi mn}^{\left(  3\right)
}\right)  $ from hypotheses (1), (2), and (4) in Sec.~\ref{sec:model}. The
explicit expressions of $K_{mn}^{\left(  3\right)  }$ and $K_{\phi
mn}^{\left(  3\right)  }$ given in Appendix.~\ref{append:d13}.

Then we seek the solutions of Eq.~(\ref{eq:ME2}) in the form of a power series
expansion in the magnitude of $\bar{H}_{pk}^{\left(  3\right)  }$, that is, a
solution of the form
\begin{equation}
s(t)=s^{(0)}\left(  t\right)  +s^{(1)}(t)+\cdots+s^{(r)}(t)+\cdots,
\label{eq:s_expansion}%
\end{equation}
for the reduced density matrix $s$ of the three-level system. Here,
$s^{\left(  0\right)  }$ is the steady state solution when no signal field is
applied to the system. However, the $r$th-order reduced density matrix
$s^{\left(  r\right)  }(t)$ is proportional to $r$th order of $\bar{H}%
_{pk}^{\left(  3\right)  }$.

In the first order approximation, we have
\begin{align}
\frac{\partial s^{\left(  0\right)  }}{\partial t} &  =\frac{1}{i\hbar}%
[H_{S}^{\left(  3\right)  },s^{\left(  0\right)  }]+\mathcal{L[}s^{\left(
0\right)  }],\label{eq:ME3_0}\\
\frac{\partial s^{\left(  1\right)  }}{\partial t} &  =\frac{1}{i\hbar}%
[H_{S}^{\left(  2\right)  },s^{\left(  1\right)  }]+\frac{1}{i\hbar}[\bar
{H}_{pk}^{\left(  3\right)  },s^{\left(  0\right)  }]+\mathcal{L[}s^{\left(
1\right)  }].\label{eq:ME3_1}%
\end{align}
The solutions of Eq.~(\ref{eq:ME3_0}) are
\begin{align}
s_{11}^{\left(  0\right)  } &  =\frac{\lambda_{21}}{\lambda_{21}y_{3}%
^{2}+\lambda_{32}y_{3}+\lambda_{21}},\\
s_{22}^{\left(  0\right)  } &  =\frac{y_{3}\lambda_{32}}{\lambda_{21}y_{3}%
^{2}+\lambda_{32}y_{3}+\lambda_{21}},\\
s_{33}^{\left(  0\right)  } &  =\frac{y_{3}^{2}\lambda_{21}}{\lambda_{21}%
y_{3}^{2}+\lambda_{32}y_{3}+\lambda_{21}},
\end{align}
all of which are $\lambda_{31}$ independent. And the other terms of
$s^{\left(  0\right)  }$ are all zeros. Having obtained $s^{\left(  0\right)
}$, we can future solve Eq.~(\ref{eq:ME3_1}). When $\bar{H}_{pk}^{\left(
3\right)  }$ takes $\bar{H}_{p5}^{\left(  3\right)  }$, we have the nonzero
matrix elements of $s^{\left(  1\right)  }$ as follows,%
\begin{align}
s_{21}^{\left(  1\right)  } &  =s_{12}^{\left(  1\right)  \ast}=\frac
{i\varepsilon_{21,5}e^{-i\omega_{5}t}\left(  s_{22}^{\left(  0\right)
}-s_{11}^{\left(  0\right)  }\right)  }{i\left(  \omega_{21}^{\left(
3\right)  }-\omega_{5}\right)  +\frac{1}{2}\Gamma_{21}^{\left(  3\right)  }%
},\label{eq:sp5_1}\\
s_{32}^{\left(  1\right)  } &  =\chi_{23}^{\left(  1\right)  \ast}%
=\frac{i\varepsilon_{32,5}^{\ast}e^{i\omega_{5}t}\left(  s_{33}^{\left(
0\right)  }-s_{22}^{\left(  0\right)  }\right)  }{i\left(  \omega
_{32}^{\left(  3\right)  }+\omega_{5}\right)  +\frac{1}{2}\Gamma_{32}^{\left(
3\right)  }}.\label{eq:sp5_2}%
\end{align}
when $\bar{H}_{pk}^{\left(  3\right)  }$ takes $\bar{H}_{p6}^{\left(
3\right)  }$, we have the nonzero matrix elements of $s^{\left(  1\right)  }$
as follows,%
\begin{align}
s_{21}^{\left(  1\right)  } &  =s_{12}^{\left(  1\right)  \ast}=\frac
{i\varepsilon_{21,6}^{\ast}e^{-i\omega_{6-}t}\left(  s_{22}^{\left(  0\right)
}-s_{11}^{\left(  0\right)  }\right)  }{-i\left(  \omega_{6-}-\omega
_{21}^{\left(  3\right)  }\right)  +\frac{1}{2}\Gamma_{21}^{\left(  3\right)
}},\label{eq:sp6_1}\\
s_{32}^{\left(  1\right)  } &  =s_{23}^{\left(  1\right)  \ast}=\frac
{i\varepsilon_{32,6}e^{i\omega_{6-}t}\left(  s_{33}^{\left(  0\right)
}-s_{22}^{\left(  0\right)  }\right)  }{i\left(  \omega_{6-}+\omega
_{32}^{\left(  3\right)  }\right)  +\frac{1}{2}\Gamma_{32}^{\left(  3\right)
}}.\label{eq:sp6_2}%
\end{align}

\subsection{Scattered current}

The scattered current at $x=0$, similarly to driving type (1), can be
represented by%
\begin{equation}
I_{s3}\left(  0,t\right)  =-\frac{iM}{2Z_{T}}\sum_{mnk}\delta_{mnk}^{\left(
3\right)  }\bar{I}_{mnk}^{\left(  3\right)  }e^{i\nu_{mnk}^{\left(  3\right)
}t}s_{nm},\label{eq:Is3}%
\end{equation}
with $\delta_{mnk}^{\left(  3\right)  }=\omega_{mn}^{\left(  3\right)  }%
+\nu_{mnk}^{\left(  3\right)  }$. Here, we have assumed that the matrix
element of $\bar{I}^{\left(  3\right)  }\left(  t\right)  $ is of the form
$\bar{I}_{mn}^{\left(  3\right)  }\left(  t\right)  =\sum_{k}\bar{I}%
_{mnk}^{\left(  3\right)  }e^{i\nu_{mnk}^{\left(  3\right)  }t}$. We hereby
also expand the scattered current as $I_{s3}=\sum_{r=0}^{\infty}$
$I_{s3}^{\left(  r\right)  }$, where $I_{s3}^{\left(  r\right)  }$ is in the
$r$th order of $\bar{H}_{pk}^{\left(  3\right)  }$. In this present paper, we
are only interested in the linear response of $\bar{H}_{pk}^{\left(  3\right)
}$, that is,%
\begin{equation}
I_{s3}^{\left(  1\right)  }\left(  0,t\right)  =-\frac{iM}{2Z_{T}}\sum
_{mnk}\delta_{mnk}^{\left(  3\right)  }\bar{I}_{mnk}^{\left(  3\right)
}e^{i\nu_{mnk}^{\left(  3\right)  }t}s_{nm}^{\left(  1\right)  }.
\end{equation}

\subsection{Probe type (5)}

When $H_{pk}^{\left(  3\right)  }$ takes $H_{p5}^{\left(  3\right)  },$ using
Eqs.~(\ref{eq:sp5_1})-(\ref{eq:sp5_2}), we have the linear response as
\begin{equation}
I_{s3}^{\left(  1\right)  }(0,t)=\operatorname{Re}\{\tilde{I}_{s3}(\omega
_{5})e^{-i\omega_{5}t}\}+\operatorname{Re}\{\tilde{I}_{s5}(\omega
_{5+})e^{-i\omega_{5+}t}\}
\end{equation}
where $\omega_{5+}=\omega_{d3}-\omega_{5+}$ is the produced difference
frequency. The amplitudes of both frequency components are respectively
$\tilde{I}_{s3}(\omega_{5})$ and $\tilde{I}_{s3}(\omega_{5+})$. The gain of
the incident current $I_{p5}$ is defined as $G_{5}=1+\tilde{I}_{s3}(\omega
_{5})/\tilde{I}_{p5}$, and the explicit expression is
\begin{align}
G_{5}= &  1+\frac{M^{2}}{2\hbar Z_{T}}\frac{\left(  s_{22}^{\left(  0\right)
}-s_{11}^{\left(  0\right)  }\right)  \lambda_{21}\cos^{2}\frac{\theta_{3}}%
{2}}{i\left(  \omega_{21}^{\left(  3\right)  }-\omega_{5}\right)  +\frac{1}%
{2}\Gamma_{21}^{\left(  3\right)  }}\nonumber\\
&  -\frac{M^{2}}{2\hbar Z_{T}}\frac{\left(  s_{33}^{\left(  0\right)  }%
-s_{22}^{\left(  0\right)  }\right)  \lambda_{21}\sin^{2}\frac{\theta_{3}}{2}%
}{i\left(  \omega_{32}^{\left(  3\right)  }+\omega_{5}\right)  +\frac{1}%
{2}\Gamma_{32}^{\left(  3\right)  }}\label{eq:G5}%
\end{align}
Meanwhile, the corresponding efficiency of frequency down conversion is
defined as $\eta_{5}=\tilde{I}_{s3}(\omega_{5+})/\tilde{I}_{p5}\sqrt
{\omega_{5}/\omega_{5+}}$ since $\left\vert \eta_{5}\right\vert ^{2}$
represents the photon number of frequency $\omega_{5+}$ produced by each
photon of frequency $\omega_{5}$ per unit time. The explicit expression of
$\eta_{5}$ is hence%
\begin{align}
\eta_{5}= &  \frac{M^{2}\sqrt{\lambda_{32}\lambda_{21}}\cos\frac{\theta_{3}%
}{2}\sin\frac{\theta_{3}}{2}}{2\hbar Z_{T}}\left[  \frac{(s_{22}^{\left(
0\right)  }-s_{11}^{\left(  0\right)  })}{i(\omega_{21}^{\left(  3\right)
}-\omega_{5})+\frac{1}{2}\Gamma_{21}^{\left(  3\right)  }}\right.  \nonumber\\
&  \left.  +\frac{(s_{33}^{\left(  0\right)  }-s_{22}^{\left(  0\right)  }%
)}{i(\omega_{32}^{\left(  3\right)  }+\omega_{5})+\frac{1}{2}\Gamma
_{32}^{\left(  3\right)  }}\right]  \exp(-i\arg\tilde{I}_{p5}).\label{eq:eta5}%
\end{align}

The two resonant points of $G_{5\,}$ and $\eta_{5}$ are respectively at
$\omega_{5}=\omega_{21}^{\left(  3\right)  }$ and $\omega_{5}=\omega
_{32}^{\left(  3\right)  }$. As we have assumed a sufficiently large
$\Omega_{31,3}$, the two points must be well separated. Therefore, we can
determine from Eq.~(\ref{eq:G5}) that the transmitted signal with frequency
$\omega_{5}$ can be both attenuated and amplified depending on the sign of
$s_{22}^{\left(  0\right)  }-s_{11}^{\left(  0\right)  }$ or $s_{33}^{\left(
0\right)  }-s_{22}^{\left(  0\right)  }$. At both points, $\left\vert
G_{5}\right\vert $ reaches the maximum or minimum gain, while $\left\vert
\eta_{5}\right\vert $ reaches the maximum conversion efficiency. To obtain the
optimal attenuation, amplification, or conversion efficiency, we can first
minimize $\Gamma_{21}^{\left(  3\right)  }$ and $\Gamma_{32}^{\left(
3\right)  }$ whose expressions are respectively
\begin{align}
\Gamma_{21}^{\left(  3\right)  } &  =\frac{M^{2}}{\hbar Z_{T}}\left(
\lambda_{21}\cos^{2}\frac{\theta_{3}}{2}+\sin^{2}\frac{\theta_{3}}{2}%
\lambda_{\text{sum}}\right)  ,\\
\Gamma_{32}^{\left(  3\right)  } &  =\frac{M^{2}}{\hbar Z_{T}}\left(  \sin
^{2}\frac{\theta_{3}}{2}\lambda_{21}+\cos^{2}\frac{\theta_{3}}{2}%
\lambda_{\text{sum}}\right)  .
\end{align}
Here, $\lambda_{\text{sum}}=\lambda_{21}+\lambda_{32}+\lambda_{31}$. The
dephasing rates $\Gamma_{21}^{\left(  3\right)  }$ and $\Gamma_{32}^{\left(
3\right)  }$ can be further reduced to
\begin{align}
\Gamma_{21}^{\left(  3\right)  } &  =\frac{M^{2}}{\hbar Z_{T}}\left[
\lambda_{21}\cos^{2}\frac{\theta_{3}}{2}+\sin^{2}\frac{\theta_{3}}{2}\left(
\lambda_{21}+\lambda_{32}\right)  \right]  ,\\
\Gamma_{32}^{\left(  3\right)  } &  =\frac{M^{2}}{\hbar Z_{T}}\left[  \left(
\lambda_{21}+\lambda_{32}\right)  \cos^{2}\frac{\theta_{3}}{2}+\sin^{2}%
\frac{\theta_{3}}{2}\lambda_{21}\right]  ,
\end{align}
in the limit that $\lambda_{1}\ll1$ and $\lambda_{2}\ll1$. We thus assume
$\lambda_{1}\ll1$ and $\lambda_{2}\ll1$ in the following discussions of
$\eta_{5}$ and $G_{5}$.

When $\omega_{5}$ is near $\omega_{21}^{\left(  3\right)  }$, the
amplification condition (i.e., $s_{22}^{\left(  0\right)  }-s_{11}^{\left(
0\right)  }>0)$ is
\begin{equation}
\lambda_{3}y_{3}>1,
\end{equation}
and the attenuation condition (i.e., $s_{22}^{\left(  0\right)  }%
-s_{11}^{\left(  0\right)  }<0)$ is
\begin{equation}
\lambda_{3}y_{3}<1.
\end{equation}

\begin{figure}[ptbh]
\includegraphics[width=0.24\textwidth, clip]{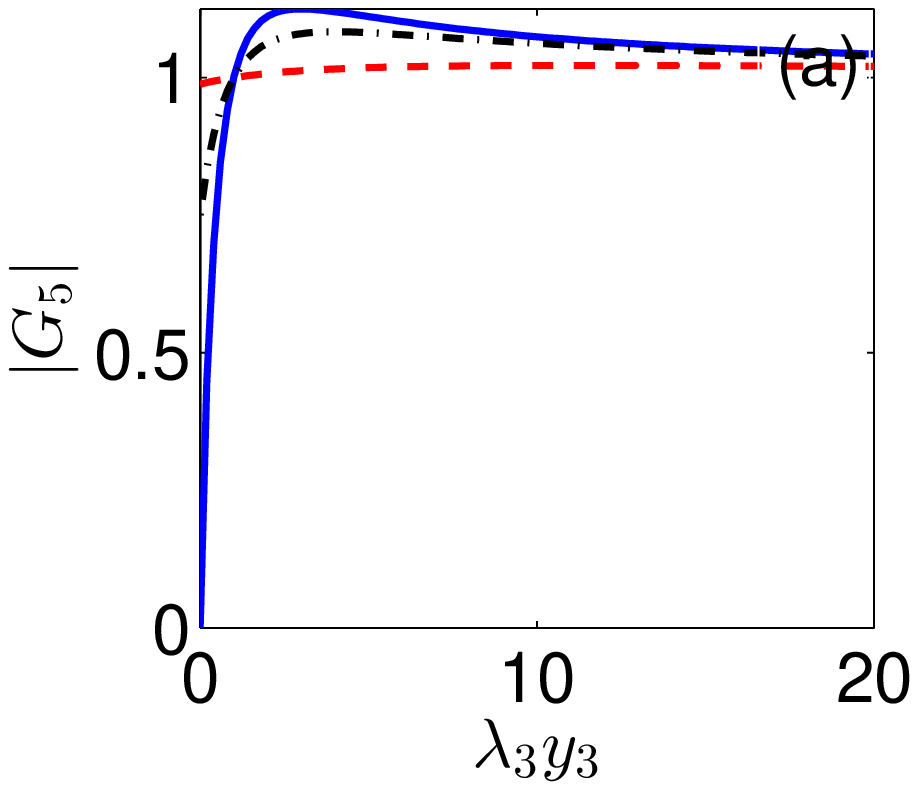}\includegraphics[width=0.24\textwidth, clip]{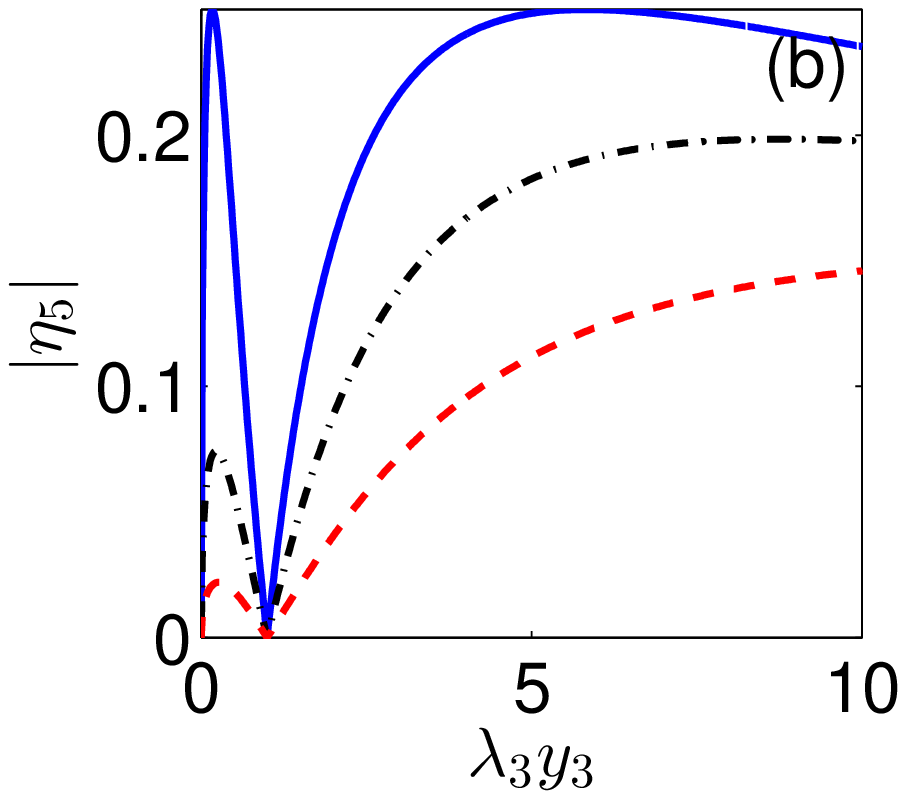}\caption{(Color
online) Probe type (5), driving type (3). The gain (a) $\left\vert
G_{5}\right\vert $ and conversion efficiency (b) $\left\vert \eta
_{5}\right\vert $ plotted as functions of $\lambda_{3}y_{3}$. Here, we have
assumed that $\omega_{5}=\omega_{21}^{\left(  3\right)  }$ and $\lambda
_{1},\lambda_{2}\ll1$. In (a), $y_{3}=0$ (solid blue), $1$ (dash-dotted
black), and $4$ (dashed red), respectively. In (b), $y_{3}=0$ (solid blue),
$1$ (dash-dotted black), and $2$ (dashed red), respectively. }%
\label{fig:probe5}%
\end{figure}

We now further seek the limitation value of $\left\vert G_{5}\right\vert $
when $\omega_{5}=\omega_{21}^{\left(  3\right)  }$. In this case, $G_{5}$ is
reduced to%
\begin{equation}
G_{5}=1+\frac{1}{1+y_{3}+\lambda_{3}y_{3}}\frac{\lambda_{3}y_{3}-1}{y_{3}%
^{2}+\lambda_{3}y_{3}+1},\label{eq:G5_p1}%
\end{equation}
with $y_{3}=\tan^{2}(\theta_{3}/2)$. In Eq.~(\ref{eq:G5_p1}), $y_{3}$ and
$\lambda_{3}y_{3}$ will be regarded as independent parameters. In the
amplification case, $|G_{5}|$ can be maximized, yielding
\begin{equation}
G_{5}=1+\frac{1}{\left(  \sqrt{A_{1}}+\sqrt{A_{2}}\right)  ^{2}}%
\label{eq:G5_p1_amp}%
\end{equation}
when the condition $\lambda_{3}y_{3}=\sqrt{A_{1}A_{2}}+1$ holds. Here,
parameters $A_{1}$ and $A_{2}$ respectively take $A_{1}=y_{3}+2$ and
$A_{2}=y_{3}^{2}+2$. Furthermore, the condition $y_{3}\ll1$ yields the optimal
amplification, yielding
\begin{equation}
G_{5}=1\frac{1}{8},
\end{equation}
where the condition for $\lambda_{3}y_{3}$ becomes $\lambda_{3}y_{3}=3$. In
the resonant driving case, using Eq.~(\ref{eq:G5_p1_amp}), the optimal $G_{5}$
for amplification can be reduced to
\begin{equation}
G_{5}=1\frac{1}{12}\text{,}%
\end{equation}
where the conditions become $y_{3}=1$ and $\lambda_{3}=4$. In the attenuation
case, $|G_{5}|$ can be minimized, yielding
\begin{equation}
G_{5}=1-\frac{1}{y_{3}^{3}+y_{3}^{2}+y_{3}+1},\label{eq:G5_p1_att}%
\end{equation}
when $\lambda_{3}y_{3}\ll1$. Furthermore, the condition $y_{3}\ll1$ yields the
optimal gain for attenuation, that is,%
\begin{equation}
G_{5}=0.
\end{equation}
In the resonant driving case, from Eq.~(\ref{eq:G5_p1_att}), the optimal
$G_{5}$ for attenuation can be reduced to
\begin{equation}
G_{5}=\frac{3}{4},
\end{equation}
where the conditions are $y_{3}=1$ and $\lambda_{3}\ll1$. In
Fig.~\ref{fig:probe5}(a), $\eta_{5}$ takes Eq.~(\ref{eq:G5_p1}). We have
plotted $\left\vert G_{5}\right\vert $ as the function of $\lambda_{3}y_{3}$
when $y_{3}$ takes 0, 1, and 4, respectively. When $\lambda_{3}y_{3}$
increases from $\lambda_{3}y_{3}<1$, to $\lambda_{3}y_{3}=1$, and then to
$\lambda_{3}y_{3}>1$, $\left\vert G_{5}\right\vert $ sequentially exhibits
attenuation ($\left\vert G_{5}\right\vert <1$), transparency ($\left\vert
G_{5}\right\vert =1$), and amplification ($\left\vert G_{5}\right\vert >1$).
Despite the regime of $\lambda_{3}y_{3}$, the increase of $y_{3}$ will always
weaken the attenuation or amplification of the probe signal.

When $\omega_{5}$ is near $-\omega_{32}^{\left(  3\right)  }$, The
amplification condition (i.e., $s_{33}^{\left(  0\right)  }-s_{22}^{\left(
0\right)  }<0$) is
\begin{equation}
\lambda_{3}y_{3}^{-1}>1,
\end{equation}
and the attenuation condition (i.e., $s_{33}^{\left(  0\right)  }%
-s_{22}^{\left(  0\right)  }>0)$ is
\begin{equation}
\lambda_{3}y_{3}^{-1}<1.
\end{equation}
We now further seek the limitation value of $\left\vert G_{5}\right\vert $
when $\omega_{5}=-\omega_{32}^{\left(  3\right)  }$. In this case, $G_{5}$ is
reduced to
\begin{equation}
G_{5}=1+\frac{1}{1+y_{3}^{-1}+\lambda_{3}y_{3}^{-1}}\frac{\lambda_{3}%
y_{3}^{-1}-1}{1+\lambda_{3}y_{3}^{-1}+y_{3}^{-2}}.\label{eq:G5_p2}%
\end{equation}
The similarity can be easily seen between Eqs.~(\ref{eq:G5_p2}) and
(\ref{eq:G5_p1}). We thus directly have that in the amplification case, the
gain $G_{5}$ for amplification can be maximized as
\begin{equation}
G_{5}=1+\frac{1}{\left(  \sqrt{B_{1}}+\sqrt{B_{2}}\right)  ^{2}}%
\label{eq:G5_p2_amp}%
\end{equation}
when $\lambda_{3}y_{3}^{-1}=\sqrt{B_{1}B_{2}}+1$ with $B_{1}=y_{3}^{-1}+2$,
$B_{2}=y_{3}^{-2}+2$. Furthermore, optimal gain for amplification can be
achieved as
\begin{equation}
G_{5}=1\frac{1}{8},
\end{equation}
when $y_{3}^{-1}\ll1$ and $\lambda_{3}y_{3}^{-1}=3$. From
Eq.~(\ref{eq:G5_p2_amp}), the optimal gain for amplification can be reduced
to
\begin{equation}
G_{5}=1\frac{1}{12},
\end{equation}
in the resonant driving case where the conditions are $y_{3}=1$, and
$\lambda_{3}=4$. In the attenuation case, the gain for attenuation can be
minimized as%
\begin{equation}
G_{5}=1-\frac{1}{y_{3}^{-3}+y_{3}^{-2}+y_{3}^{-1}+1},\label{eq:G5_p2_att}%
\end{equation}
when $\lambda_{3}y_{3}^{-1}\ll1$. Furthermore, the condition $y_{3}^{-1}\ll1$
yields the optimal gain for attenuation, that is,%
\begin{equation}
G_{5}=0.
\end{equation}
From Eq.~(\ref{eq:G5_p2_att}), the optimal $G_{5}$ for attenuation can be
reduced to%
\begin{equation}
G_{5}=\frac{3}{4},
\end{equation}
in the resonant driving case where the condition are $y_{3}=1$ and
$\lambda_{3}\ll1$. The behaviours of $\left\vert G_{5}\right\vert $ can be
similarly explained through Fig.~\ref{fig:probe5}(a) when $G_{5}$ takes
Eq.~(\ref{eq:G5_p2}).

We now seek the limitation of $|\eta_{5}|$ when $\omega_{5}=\omega
_{21}^{\left(  3\right)  }$. In this case, the conversion efficiency $\eta
_{5}$ is reduced to%
\begin{equation}
\eta_{5}=\frac{\exp(-i\arg\tilde{I}_{p5})\sqrt{\lambda_{3}y_{3}}}%
{1+y_{3}+\lambda_{3}y_{3}}\frac{\lambda_{3}y_{3}-1}{1+\lambda_{3}y_{3}%
+y_{3}^{2}}.\label{eq:eta5_p1}%
\end{equation}
Here, $y_{3}$ and $\lambda_{3}y_{3}$ will be continuously treated as
independent. When $y_{3}\ll1$, the conversion efficiency $\eta_{5}$ can be
maximized as
\begin{equation}
\eta_{5}=\frac{\exp(-i\arg\tilde{I}_{p5})\sqrt{\lambda_{3}y_{3}}}%
{1+\lambda_{3}y_{3}}\frac{\lambda_{3}y_{3}-1}{1+\lambda_{3}y_{3}}.
\end{equation}
Furthermore, the condition $\lambda_{3}y_{3}=3\pm2\sqrt{2}$ yields the optimal
conversion efficiency as%
\begin{equation}
\eta_{5}=\pm\frac{1}{4}\exp(-i\arg\tilde{I}_{p5}).
\end{equation}
In the resonant driving case, $y_{3}=1$, and the optimal conversion efficiency
reads%
\begin{equation}
\eta_{5}=0.198\,38\exp(-i\arg\tilde{I}_{p5}).
\end{equation}
under the condition $\lambda_{3}=\left(  9+\sqrt{73}\right)  /2$. In
Fig.~\ref{fig:probe5}(b), $\eta_{5}$ takes Eq.~(\ref{eq:eta5_p1}). We have
plotted $\left\vert \eta_{5}\right\vert $ as the functions of $\lambda
_{3}y_{3}$ when $y_{3}$ takes 0, 1, and 2, respectively. When $\lambda
_{3}y_{3}=1$, we can see $\left\vert \eta_{5}\right\vert =0$, indicating the
switch off of conversion process. Besides this point, there are two peaks.
When $y_{3}=0$, the two peaks are of the same value. When $y_{3}$ takes 1 and
2, the peak at the right takes the largest value. Besides, the larger value
$y_{3}$ takes, the smaller $\left\vert \eta_{5}\right\vert $ becomes. It is a
similar case when $\omega_{5+}=-\omega_{32}^{\left(  3\right)  }$, where the
conversion efficiency becomes%
\begin{equation}
\eta_{5}=\frac{\left(  1-\lambda_{3}y_{3}^{-1}\right)  \sqrt{\lambda_{3}%
y_{3}^{-1}}\exp(-i\arg\tilde{I}_{p5})}{\left(  1+\lambda_{3}y_{3}^{-1}%
+y_{3}^{-2}\right)  \left(  \lambda_{3}y_{3}^{-1}+y_{3}^{-1}+1\right)
}.\label{eq:eta5_p2}%
\end{equation}
The similarity can be easily found between Eqs.~(\ref{eq:eta5_p2}) and
(\ref{eq:eta5_p1}). Thus, it can be directly given that at $y_{3}^{-1}\ll1$
and $\lambda_{3}y_{3}^{-1}=3\pm2\sqrt{2}$, the optimal conversion efficiency
reads
\begin{equation}
\eta_{5}=\mp\frac{1}{4}\exp(-i\arg\tilde{I}_{p5}).
\end{equation}
In the resonant driving case, $y_{3}=1$, and the optimal conversion efficiency
reads
\begin{equation}
\eta_{5}=-0.198\,38
\end{equation}
when $\lambda_{3}=\left(  9+\sqrt{73}\right)  /2$. When $\eta_{5}$
takes\ Eq.~(\ref{eq:eta5_p2}), the behavious of $\left\vert \eta
_{5}\right\vert $ can also be explained though Fig.~\ref{fig:probe5}(b).

\subsection{Probe type (6)}

When $H_{pk}^{\left(  2\right)  }$ takes $H_{p6}^{\left(  2\right)  },$ using
Eqs.~(\ref{eq:sp6_1})-(\ref{eq:sp6_2}), we have the linear response as
\begin{equation}
I_{s3}^{\left(  1\right)  }(0,t)=\operatorname{Re}\{\tilde{I}_{s3}(\omega
_{6})e^{-i\omega_{6}t}\}+\operatorname{Re}\{\tilde{I}_{s3}(\omega
_{6-})e^{-i\omega_{6-}t}\}.
\end{equation}
The amplitudes of both frequency components are respectively $\tilde{I}%
_{s3}\left(  \omega_{6}\right)  $ and $\tilde{I}_{s3}\left(  \omega
_{6-}\right)  $. The gain of the incident current $I_{p6}$ is defined as
$G_{6}=1+\tilde{I}_{s3}\left(  \omega_{6}\right)  /\tilde{I}_{p6}$, and the
explicit expression is
\begin{align}
G_{6}= &  1-\frac{M^{2}}{2\hbar Z_{T}}\frac{\lambda_{32}\sin^{2}\frac
{\theta_{3}}{2}\left(  s_{22}^{\left(  0\right)  }-s_{11}^{\left(  0\right)
}\right)  }{i\left(  \omega_{6-}-\omega_{21}^{\left(  3\right)  }\right)
+\frac{1}{2}\Gamma_{21}^{\left(  3\right)  }}\nonumber\\
&  +\frac{M^{2}}{2\hbar Z_{T}}\frac{\lambda_{32}\cos^{2}\frac{\theta_{3}}%
{2}\left(  s_{33}^{\left(  0\right)  }-s_{22}^{\left(  0\right)  }\right)
}{i\left(  \omega_{6-}+\omega_{32}^{\left(  3\right)  }\right)  +\frac{1}%
{2}\Gamma_{32}^{\left(  3\right)  }}.\label{eq:G6}%
\end{align}
Meanwhile, the corresponding efficiency of frequency down conversion is
defined as $\eta_{6}=\tilde{I}\left(  \omega_{6-}\right)  /\tilde{I}_{p6}%
\sqrt{\omega_{6}/\omega_{6-}}$ since $\left\vert \eta_{6}\right\vert ^{2}$
represents the photon number of frequency $\omega_{6-}$ produced by each
photon of frequency $\omega_{6}$ per unit time. The explicit expression of
$\eta_{6}$ is hence%
\begin{align}
\eta_{6}= &  \frac{M^{2}\sqrt{\lambda_{21}\lambda_{32}}\sin\frac{\theta_{3}%
}{2}\cos\frac{\theta_{3}}{2}}{-2\hbar Z_{T}}\left[  \frac{(s_{22}^{\left(
0\right)  }-s_{11}^{\left(  0\right)  })}{-i(\omega_{6-}-\omega_{21}^{\left(
3\right)  })+\frac{1}{2}\Gamma_{21}^{\left(  3\right)  }}\right.  \nonumber\\
&  \left.  +\frac{(s_{33}^{\left(  0\right)  }-s_{22}^{\left(  0\right)  }%
)}{-i(\omega_{6-}+\omega_{32}^{\left(  3\right)  })+\frac{1}{2}\Gamma
_{32}^{\left(  3\right)  }}\right]  \exp(-i\arg\tilde{I}_{p6}).\label{eq:eta6}%
\end{align}

The two resonant points of $G_{6\,}$ and $\eta_{6}$ are respectively at
$\omega_{6-}=\omega_{21}^{\left(  3\right)  }$ and $\omega_{6-}=-\omega
_{32}^{\left(  3\right)  }$. As we have assumed a sufficiently large
$\Omega_{31,3}$, the two points must be well separated. Therefore, we can
determine from Eq.~(\ref{eq:G2}) that the transmitted signal with frequency
$\omega_{6}$ can be both attenuated or amplified. At both points, $\left\vert
G_{6}\right\vert $ reaches the maximum gain and maximum conversion efficiency,
while $\left\vert \eta_{6}\right\vert $ reaches the maximum conversion
efficiency. We will similarly assume $\lambda_{1}\ll1$ and $\lambda_{2}\ll1$
as in probe type (5) in the following discussions of $\eta_{6}$ and $G_{6}$.

\begin{figure}[ptbh]
\includegraphics[width=0.24\textwidth, clip]{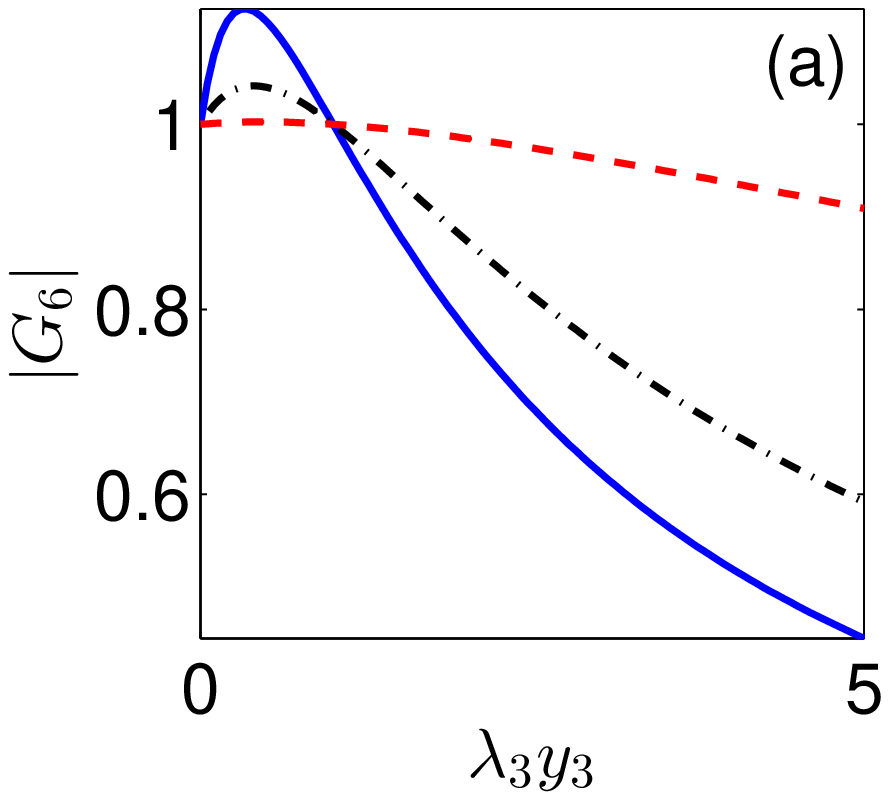}\includegraphics[width=0.24\textwidth, clip]{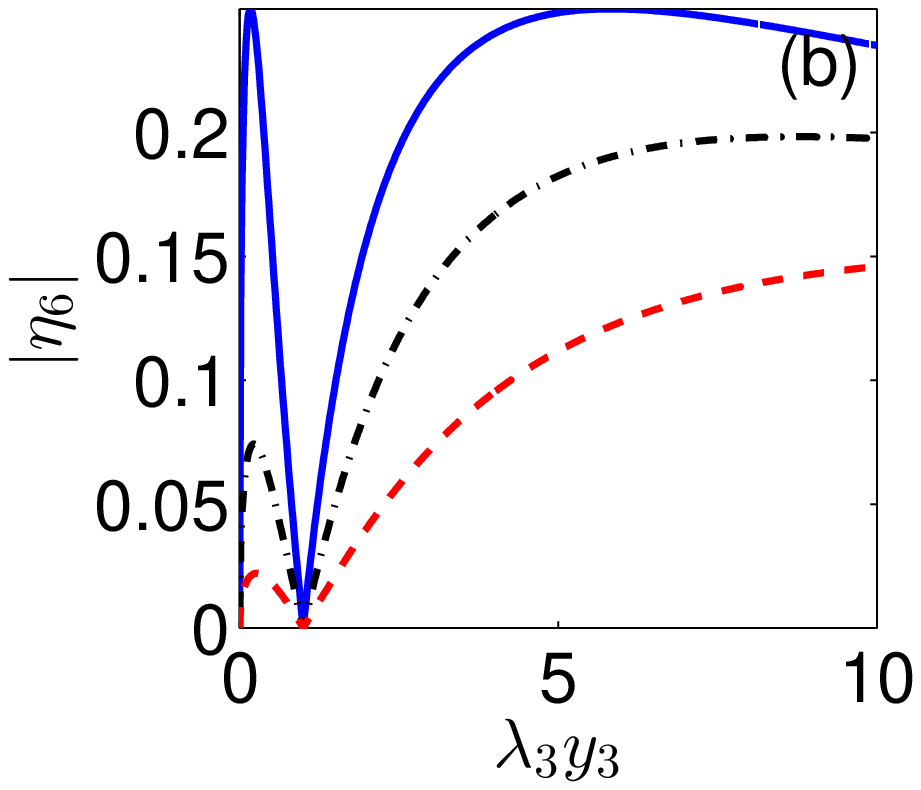}\caption{(Color
online) Probe type (6), driving type (3). The gain (a) $\left\vert
G_{6}\right\vert $ and conversion efficiency (b) $\left\vert \eta
_{6}\right\vert $ plotted as functions of $\lambda_{3}y_{3}$. Here, we have
assumed that $\omega_{6-}=\omega_{21}^{\left(  3\right)  }$ and $\lambda
_{1},\lambda_{2}\ll1$. In (a), $y_{3}=0$ (solid blue), $1$ (dash-dotted
black), and $4$ (dashed red), respectively. In (b), $y_{3}=0$ (solid blue),
$1$ (dash-dotted black), and $2$ (dashed red), respectively. }%
\label{fig:probe6}%
\end{figure}

We now further seek the limitation value of $\left\vert G_{6}\right\vert $
when $\omega_{6-}=\omega_{21}^{\left(  3\right)  }$. In this case, $G_{6}$ is
reduced to%
\begin{equation}
G_{6}=1-\frac{y_{3}\lambda_{3}}{1+y_{3}+y_{3}\lambda_{3}}\frac{y_{3}%
\lambda_{3}-1}{1+y_{3}\lambda_{3}+y_{3}^{2}}.\label{eq:G6_p1}%
\end{equation}
In the amplification case,\ the gain can be maximized as%
\begin{equation}
G_{6}=1+\frac{1}{\left(  \sqrt{A_{1}A_{4}}+\sqrt{A_{2}A_{3}}\right)  ^{2}%
},\label{eq:G6_p1_amp}%
\end{equation}
when $\lambda_{3}y_{3}=(\sqrt{A_{1}A_{2}/A_{3}A_{4}}+1)^{-1}$ with
$A_{3}=y_{3}+1$, and $A_{4}=y_{3}^{2}+1$. Furthermore, the condition
$y_{3}\ll1\ $yields the optimal gain for amplification, i.e.,%
\begin{equation}
G_{6}=1\frac{1}{8},
\end{equation}
where the condition for $\lambda_{3}y_{3}$ becomes $\lambda_{3}y_{3}=1/3$. From
Eq.~(\ref{eq:G6_p1_amp}), the optimal gain for amplification can be reduced
to
\begin{equation}
G_{6}=1\frac{1}{24},
\end{equation}
in the resonant case, where the conditions are $y_{3}=1$ and $\lambda_{3}%
=2/5$. In the attenuation case, the condition $\lambda_{3}y_{3}\gg1$ yields
the optimal gain for attenuation
\begin{equation}
G_{6}=0.
\end{equation}
In the resonant driving case, the conditions become $y_{3}=1$ and $\lambda
_{3}\gg1$ where the optimal gain for attenuation is still $G_{6}=0$. In
Fig.~\ref{fig:probe6}, $G_{6}$ takes Eq.~(\ref{eq:G6_p1}). We have plotted
$\left\vert G_{6}\right\vert $ as the function of $\lambda_{3}y_{3}$ when
$y_{3}$ takes 0, 1, and 4, respectively. When $\lambda_{3}y_{3}$ changes from
$\lambda_{3}y_{3}<1$, to $\lambda_{3}y_{3}=1$, and then to $\lambda_{3}%
y_{3}>1$, $\left\vert G_{6}\right\vert $ sequentially exhibits amplification
($\left\vert G_{6}\right\vert >1$), transparency ($\left\vert G_{6}\right\vert
=1$), and attenuation ($\left\vert G_{6}\right\vert <1$). Despite the regime
of $\lambda_{3}y_{3}$, the increase of $y_{3}$ will always weaken the
attenuation or amplification of the probe signal. It is a similar case when
$\omega_{6-}=-\omega_{32}^{\left(  3\right)  }$, where the gain becomes%
\begin{equation}
G_{6}=1+\frac{\lambda_{3}y_{3}^{-1}}{1+y_{3}^{-1}+\lambda_{3}y_{3}^{-1}}%
\frac{1-\lambda_{3}y_{3}^{-1}}{1+\lambda_{3}y_{3}^{-1}+y_{3}^{-2}%
}.\label{eq:G6_p2}%
\end{equation}
The similarity can be easily found between Eqs.~.(\ref{eq:G6_p2}) and
(\ref{eq:G6_p1}). Thus, we directly have that the gain for amplification can
be optimized as%
\begin{equation}
G_{6}=1+\frac{1}{\left(  \sqrt{B_{1}B_{4}}+\sqrt{B_{2}B_{3}}\right)  ^{2}%
},\label{eq:G6_p2_amp}%
\end{equation}
when $\lambda_{3}y_{3}^{-1}=\left(  \sqrt{B_{1}B_{2}/B_{3}B_{4}}+1\right)
^{-1}$ with $B_{1}=y_{3}^{-1}+2$, $B_{2}=y_{3}^{-2}+2$, $B_{3}=1+y_{3}^{-1}$,
and $B_{4}=1+y_{3}^{-2}$. Furthermore, the condition $y_{3}^{-1}\ll1$ yields
the optimal gain for amplification%
\begin{equation}
G_{6}=1\frac{1}{8},
\end{equation}
where $\lambda_{3}y_{3}^{-1}=1/3.$ From Eq.~(\ref{eq:G6_p2_amp}), the gain for
amplification can be reduced to
\begin{equation}
G_{6}=1\frac{1}{24},
\end{equation}
with the condition $y_{3}=1$ and $\lambda_{3}=2/5$. In the attenuation case,
the condition $\lambda_{3}y_{3}^{-1}\gg1$ yields the optimal gain as%
\begin{equation}
G_{6}=0.
\end{equation}
In the resonant case, the conditions become $y_{3}=1$ and $\lambda_{3}\gg1$
where the optimal gain for attenuation is still $G_{6}=0$. When $G_{6}$ takes
Eq.~(\ref{eq:G6_p2}), the behaviour of $\left\vert G_{6}\right\vert $ can also
be investigated through Fig.~\ref{eq:G6_p2}(a).

We now seek the limitation of $|\eta_{6}|$ when $\omega_{6-}=\omega
_{21}^{\left(  3\right)  }$. In this case, $\eta_{6}$ is reduced to%
\begin{equation}
\eta_{6}=-\frac{\exp(-i\arg\tilde{I}_{p6})\sqrt{\lambda_{3}y_{3}}}%
{1+y_{3}+\lambda_{3}y_{3}}\frac{\lambda_{3}y_{3}-1}{1+\lambda_{3}y_{3}%
+y_{3}^{2}}. \label{eq:eta6_p1}%
\end{equation}
We find Eqs.~(\ref{eq:eta6_p1}) and (\ref{eq:eta5_p1}) are equivalent except
for a global constant. We thus directly have the optimal conversion efficiency%
\begin{equation}
\eta_{6}=\mp\frac{1}{4}\exp\left(  -i\arg\tilde{I}_{p6}\right)  .
\end{equation}
when $y_{3}\ll1$ and $\lambda_{3}y_{3}=3\pm2\sqrt{2}$. In the resonant driving
case, $y_{3}=1$, and the optimal conversion efficiency becomes
\begin{equation}
\eta_{6}=-0.198\,38\exp\left(  -i\arg\tilde{I}_{p6}\right)  ,
\end{equation}
in the condition that $\lambda_{3}=\left(  9+\sqrt{73}\right)  /2$. It is a
similar case when $\omega_{6-}=-\omega_{32}^{\left(  3\right)  }$, where the
conversion efficiency becomes%
\begin{equation}
\eta_{6}=-\frac{\exp(-i\arg\tilde{I}_{p6})\sqrt{\lambda_{3}y_{3}^{-1}}%
}{1+y_{3}^{-1}+\lambda_{3}y_{3}^{-1}}\frac{1-\lambda_{3}y_{3}^{-1}}%
{1+\lambda_{3}y_{3}^{-1}+y_{3}^{-2}}. \label{eq:eta6_p2}%
\end{equation}
With the condition $y_{3}^{-1}\ll1$ and $\lambda_{3}y_{3}^{-1}=3\pm2\sqrt{2}$,
the optimal conversion efficiency reaches%
\begin{equation}
\eta_{6}=\pm\frac{1}{4}\exp\left(  -i\arg\tilde{I}_{p6}\right)  .
\end{equation}
In the resonant driving case, $y_{3}=1$, and we can achieve the optimal
conversion efficiency as
\begin{equation}
\eta_{6}=0.198\,38\exp\left(  -i\arg\tilde{I}_{p6}\right)  ,
\end{equation}
when $\lambda_{3}=\left(  9+\sqrt{73}\right)  /2$. For completeness, we also
plot Fig.~\ref{fig:probe6}(b). Both Eqs.~(\ref{eq:eta6_p1}) and
(\ref{eq:eta6_p2}) can be described by Fig.~\ref{fig:probe6}(b).

\section{Conclusions and Discussions\label{sec:discussion}}

In summary, using a three-level three-junction flux qubit circuit as an
example, we study how the frequency conversion and signal amplification (or
attenuation) can be realized when the inversion symmetry of the potential
energy is broken. We mention that the microwave amplification has recently
been experimentally \color{blue}realized~\cite{Olig,nec2010science}
\color{black} in a three-level system constructed by four-junction flux qubit
circuits. However, our study provides a full picture for understanding the
microwave frequency conversion and amplification (or attenuation) in the case
of all possible driving and probing. As a summary, we list in
Table.~\ref{tab:LimSummary} the maximum (or minimum) gains and maximum
conversion efficiencies at different driving and probe types. The conditions
for achieving these values are also appended in this table.

Based on different configurations of the applied driving and probing fields,
we classify our study into three types. We find that a single three-level
superconducting flux qubit circuit is enough to complete the microwave
frequency conversion, amplification (or attenuation) of weak signal fields. In
particular, we find, (i) in the driving types (1) and (2), the three-level
system can convert the driving and signal fields into the ones with new
frequencies, which we call down conversion or up conversion, respectively. Due
to the energy loss in the reflection and conversion, the incident signal field
suffers the attenuation after transmitted in these two driving types; (ii)
however, both amplification and attenuation can occur in the driving type (3),
whether the amplification or the attenuation depends on the parameter
condition; (iii) given a definite flux bias, when the driving and signal
detunings are adjusted properly, the maximum conversion efficiencies and gains
nearly do not depend on the driving strength.%

%TCIMACRO{\TeXButton{B}{\begin{widetext}
%\begingroup\squeezetable\begin{table*}[tbp] \centering}}%
%BeginExpansion
\begin{widetext}
\begingroup\squeezetable\begin{table*}[tbp] \centering
%EndExpansion
\caption{Maximum or minimum gains and maximum conversion efficiencies with the
corresponding conditions for achieving them. In $\left\vert G_{k}\right\vert
$ column, $\left\vert G_{k}\right\vert <1$ means the attenuation is
optimized while $\left\vert G_{k}\right\vert >1$ means the amplification is
optimized. Here, $l$ denotes the driving type and $k$ denotes the probe type.$\ $Additionally, $C_{d}$ represents the condition relating to the driving
type and $C_{p}$ represents the condition relating to the probe type. In
the detuned (resonant) driving case, the maximum or minimum $|G_{k}|$ is
achieved at the condition $C_{d}\wedge C_{p}\wedge C_{\text{Detuned
(Resonant)}}^{G}$ while the maximum $|\eta _{k}|$ is achieved at the
condition $C_{d}\wedge C_{p}\wedge C_{\text{Detuned (Resonant)}}^{\eta }$. The symbol $A_{14}=\sqrt{A_{1}A_{2}/A_{3}A_{4}}$ and $B_{14}=\sqrt{B_{1}B_{2}/B_{3}B_{4}}.$}%
\begin{tabular}
[c]{cc|cc|cc|cc|cc|cc}\hline\hline
&  &  &  & \multicolumn{2}{|c}{Detuned driving} & \multicolumn{2}{|c}{Resonant
driving} & \multicolumn{2}{|c}{Detuned driving} & \multicolumn{2}{|c}{Resonant
driving}\\\cline{5-12}%
$l$ & $C_{d}$ & $k$ & $C_{p}$ & $C_{\text{Detuned}}^{G}$ & $\left\vert
G_{k}\right\vert $ & $C_{\text{Resonant}}^{G}$ & $\left\vert G_{k}\right\vert
$ & $C_{\text{Detuned}}^{\eta}$ & $\left\vert \eta_{k}\right\vert $ &
$C_{\text{Resonant}}^{\eta}$ & $\left\vert \eta_{k}\right\vert $\\\hline
1 & $%
\begin{array}
[c]{c}%
\lambda_{2}\gg1\\
\lambda_{3}\gg1
\end{array}
$ & 1 & $\omega_{1}=\omega_{31}^{\left(  1\right)  }$ & $%
\begin{array}
[c]{c}%
\lambda_{1}\gg1\\
y_{1}\ll1
\end{array}
$ & 0 & $%
\begin{array}
[c]{c}%
\lambda_{1}\gg1\\
y_{1}=1
\end{array}
$ & $\frac{3}{4}$ & $%
\begin{array}
[c]{c}%
\lambda_{1}=1\\
y_{1}=0.36349
\end{array}
$ & $0.19529$ & $%
\begin{array}
[c]{c}%
\lambda_{1}=1\\
y_{1}=1
\end{array}
$ & $\frac{1}{8}$\\\hline
1 & $%
\begin{array}
[c]{c}%
\lambda_{2}\gg1\\
\lambda_{3}\gg1
\end{array}
$ & 1 & $\omega_{1}=\omega_{32}^{\left(  1\right)  }$ & $%
\begin{array}
[c]{c}%
\lambda_{1}\gg1\\
y_{1}^{-1}\ll1
\end{array}
$ & 0 & $%
\begin{array}
[c]{c}%
\lambda_{1}\gg1\\
y_{1}=1
\end{array}
$ & $\frac{3}{4}$ & $%
\begin{array}
[c]{c}%
\lambda_{1}=1\\
y_{1}^{-1}=0.36349
\end{array}
$ & $0.19529$ & $%
\begin{array}
[c]{c}%
\lambda_{1}=1\\
y_{1}=1
\end{array}
$ & $\frac{1}{8}$\\\hline
1 & $%
\begin{array}
[c]{c}%
\lambda_{2}\gg1\\
\lambda_{3}\gg1
\end{array}
$ & 2 & $\omega_{2+}=\omega_{31}^{\left(  1\right)  }$ & $%
\begin{array}
[c]{c}%
\lambda_{1}\ll1\\
y_{1}=0.6573
\end{array}
$ & $0.72305$ & $%
\begin{array}
[c]{c}%
\lambda_{1}\ll1\\
y_{1}=1
\end{array}
$ & $\frac{3}{4}$ & $%
\begin{array}
[c]{c}%
\lambda_{1}=1\\
y_{1}=0.36349
\end{array}
$ & $0.19529$ & $%
\begin{array}
[c]{c}%
\lambda_{1}=1\\
y_{1}=1
\end{array}
$ & $\frac{1}{8}$\\\hline
1 & $%
\begin{array}
[c]{c}%
\lambda_{2}\gg1\\
\lambda_{3}\gg1
\end{array}
$ & 2 & $\omega_{2+}=\omega_{32}^{\left(  1\right)  }$ & $%
\begin{array}
[c]{c}%
\lambda_{1}\ll1\\
y_{1}^{-1}=0.6573
\end{array}
$ & $0.72305$ & $%
\begin{array}
[c]{c}%
\lambda_{1}\ll1\\
y_{1}=1
\end{array}
,$ & $\frac{3}{4}$ & $%
\begin{array}
[c]{c}%
\lambda_{1}=1\\
y_{1}^{-1}=0.36349
\end{array}
$ & $0.19529$ & $%
\begin{array}
[c]{c}%
\lambda_{1}=1\\
y_{1}=1
\end{array}
$ & $\frac{1}{8}$\\\hline
2 & $%
\begin{array}
[c]{c}%
\lambda_{1}\gg1\\
\lambda_{3}\ll1
\end{array}
$ & 3 & $\omega_{3-}=\omega_{21}^{\left(  2\right)  }$ & $\lambda_{2}y_{2}%
\gg1$ & 0 & $%
\begin{array}
[c]{c}%
\lambda_{2}\gg1\\
y_{2}=1
\end{array}
$ & 0 & $\lambda_{2}y_{2}=1$ & $\frac{1}{2}$ & $%
\begin{array}
[c]{c}%
\lambda_{2}=1\\
y_{2}=1
\end{array}
$ & $\frac{1}{2}$\\\hline
2 & $%
\begin{array}
[c]{c}%
\lambda_{1}\gg1\\
\lambda_{3}\ll1
\end{array}
$ & 3 & $\omega_{3-}=\omega_{31}^{\left(  2\right)  }$ & $\lambda_{2}%
y_{2}^{-1}\gg1$ & 0 & $%
\begin{array}
[c]{c}%
\lambda_{2}\gg1\\
y_{2}=1
\end{array}
$ & 0 & $\lambda_{2}y_{2}^{-1}=1$ & $\frac{1}{2}$ & $%
\begin{array}
[c]{c}%
\lambda_{2}=1\\
y_{2}=1
\end{array}
$ & $\frac{1}{2}$\\\hline
2 & $%
\begin{array}
[c]{c}%
\lambda_{1}\gg1\\
\lambda_{3}\ll1
\end{array}
$ & 4 & $\omega_{4}=\omega_{21}^{\left(  2\right)  }$ & $\lambda_{2}y_{2}\ll1$
& 0 & $%
\begin{array}
[c]{c}%
\lambda_{2}\ll1\\
y_{2}=1
\end{array}
$ & 0 & $\lambda_{2}y_{2}=1$ & $\frac{1}{2}$ & $%
\begin{array}
[c]{c}%
\lambda_{2}=1\\
y_{2}=1
\end{array}
,$ & $\frac{1}{2}$\\\hline
2 & $%
\begin{array}
[c]{c}%
\lambda_{1}\gg1\\
\lambda_{3}\ll1
\end{array}
$ & 4 & $\omega_{4}=\omega_{31}^{\left(  2\right)  }$ & $\lambda_{2}y_{2}%
^{-1}\ll1$ & 0 & $%
\begin{array}
[c]{c}%
\lambda_{2}\ll1\\
y_{2}=1
\end{array}
$ & 0 & $\lambda_{2}y_{2}^{-1}=1$ & $\frac{1}{2}$ & $%
\begin{array}
[c]{c}%
\lambda_{2}=1\\
y_{2}=1
\end{array}
,$ & $\frac{1}{2}$\\\hline
3 & $%
\begin{array}
[c]{c}%
\lambda_{1}\ll1\\
\lambda_{2}\ll1
\end{array}
$ & 5 & $\omega_{5}=\omega_{21}^{\left(  3\right)  }$ & $%
\begin{array}
[c]{c}%
\lambda_{3}y_{3}=3\\
y_{3}\ll1
\end{array}
$ & $1\frac{1}{8}$ & $%
\begin{array}
[c]{c}%
\lambda_{3}=4\\
y_{3}=1
\end{array}
$ & $1\frac{1}{12}$ & $%
\begin{array}
[c]{c}%
\lambda_{3}y_{3}=3\pm2\sqrt{2}\\
y_{3}\ll1
\end{array}
$ & $\frac{1}{4}$ & $%
\begin{array}
[c]{c}%
\lambda_{3}=(9+\sqrt{73})/2\\
y_{3}=1
\end{array}
$ & $0.19838$\\\hline
3 & $%
\begin{array}
[c]{c}%
\lambda_{1}\ll1\\
\lambda_{2}\ll1
\end{array}
$ & 5 & $\omega_{5}=\omega_{21}^{\left(  3\right)  }$ & $%
\begin{array}
[c]{c}%
y_{3}\lambda_{3}\ll1\\
y_{3}\ll1
\end{array}
$ & 0 & $%
\begin{array}
[c]{c}%
\lambda_{3}\ll1\\
y_{3}=1
\end{array}
$ & $\frac{3}{4}$ & - & - & - & -\\\hline
3 & $%
\begin{array}
[c]{c}%
\lambda_{1}\ll1\\
\lambda_{2}\ll1
\end{array}
$ & 5 & $\omega_{5}=-\omega_{32}^{\left(  3\right)  }$ & $%
\begin{array}
[c]{c}%
\lambda_{3}y_{3}^{-1}=3\\
y_{3}^{-1}\ll1
\end{array}
$ & $1\frac{1}{8}$ & $%
\begin{array}
[c]{c}%
\lambda_{3}=4\\
y_{3}=1
\end{array}
$ & $1\frac{1}{12}$ & $%
\begin{array}
[c]{c}%
\lambda_{3}y_{3}^{-1}=3\pm2\sqrt{2}\\
y_{3}^{-1}\ll1
\end{array}
$ & $\frac{1}{4}$ & $%
\begin{array}
[c]{c}%
\lambda_{3}=(9+\sqrt{73})/2\\
y_{3}=1
\end{array}
$ & $0.19838$\\\hline
3 & $%
\begin{array}
[c]{c}%
\lambda_{1}\ll1\\
\lambda_{2}\ll1
\end{array}
$ & 5 & $\omega_{5}=-\omega_{32}^{\left(  3\right)  }$ & $%
\begin{array}
[c]{c}%
\lambda_{3}y_{3}^{-1}\ll1\\
y_{3}^{-1}\ll1
\end{array}
$ & $0$ & $%
\begin{array}
[c]{c}%
\lambda_{3}\ll1\\
y_{3}\ll1
\end{array}
$ & $\frac{3}{4}$ & - & - & - & -\\\hline
3 & $%
\begin{array}
[c]{c}%
\lambda_{1}\ll1\\
\lambda_{2}\ll1
\end{array}
$ & 6 & $\omega_{6-}=\omega_{21}^{\left(  3\right)  }$ & $%
\begin{array}
[c]{c}%
\lambda_{3}y_{3}=1/3\\
y_{3}\ll1
\end{array}
$ & $1\frac{1}{8}$ & $%
\begin{array}
[c]{c}%
\lambda_{3}=\frac{2}{5}\\
y_{3}=1
\end{array}
$ & $1\frac{1}{24}$ & $,%
\begin{array}
[c]{c}%
\lambda_{3}y_{3}=3\pm2\sqrt{2}\\
y_{3}\ll1
\end{array}
$ & $\frac{1}{4}$ & $%
\begin{array}
[c]{c}%
\lambda_{3}=(9+\sqrt{73})/2\\
y_{3}=1
\end{array}
$ & $0.19838$\\\hline
3 & $%
\begin{array}
[c]{c}%
\lambda_{1}\ll1\\
\lambda_{2}\ll1
\end{array}
$ & 6 & $\omega_{6-}=\omega_{21}^{\left(  3\right)  }$ & $\lambda_{3}y_{3}%
\gg1$ & 0 & $%
\begin{array}
[c]{c}%
\lambda_{3}\gg1\\
y_{3}=1
\end{array}
$ & 0 &  &  &  & \\\hline
3 & $%
\begin{array}
[c]{c}%
\lambda_{1}\ll1\\
\lambda_{2}\ll1
\end{array}
$ & 6 & $\omega_{6-}=-\omega_{32}^{\left(  3\right)  }$ & $%
\begin{array}
[c]{c}%
\lambda_{3}y_{3}^{-1}=1/3\\
y_{3}^{-1}\ll1
\end{array}
$ & $1\frac{1}{8}$ & $%
\begin{array}
[c]{c}%
\lambda_{3}=\frac{2}{5}\\
y_{3}=1
\end{array}
$ & $1\frac{1}{24}$ & $%
\begin{array}
[c]{c}%
\lambda_{3}y_{3}^{-1}=3\pm2\sqrt{2}\\
y_{3}^{-1}\ll1
\end{array}
$ & $\frac{1}{4}$ & $%
\begin{array}
[c]{c}%
\lambda_{3}=(9+\sqrt{73})/2\\
y_{3}=1
\end{array}
$ & $0.19838$\\\hline
3 & $%
\begin{array}
[c]{c}%
\lambda_{1}\ll1\\
\lambda_{2}\ll1
\end{array}
$ & 6 & $\omega_{6-}=-\omega_{32}^{\left(  3\right)  }$ & $\lambda_{3}%
y_{3}^{-1}\gg1$ & 0 & $%
\begin{array}
[c]{c}%
\lambda_{3}\gg1\\
y_{3}=1
\end{array}
$ & 0 &  &  &  & \\\hline\hline
\end{tabular}
\label{tab:LimSummary}%
%TCIMACRO{\TeXButton{E}{\end{table*}
%\endgroup\end{widetext}}}%
%BeginExpansion
\end{table*}
\endgroup\end{widetext}%
%EndExpansion

Although our study focuses on a three-level superconducting flux qubit
circuit, the method used here can be easily applied to a superconducting
phase~\cite{Martinis,Hakonen1,Hakonen2} and Xmon~\cite{Xmon} qubit circuits or
other quantum circuits in which the inversion symmetry of their potential
energy is broken. In contrast to large anharmonicity of the flux qubit
circuits, the superconducting phase and Xmon qubit circuits have small
anharmonicity. Therefore the information leakage should be more carefully
studied when these processes are demonstrated. We note that the transmon qubit
circuits~\cite{transmon1,transmon} have well defined symmetry when the
effective offset charge is at the optimal point, thus the transmon qubit
circuit for its three lowest energy levels has ladder-type transitions, and
the three-wave mixing cannot be realized. However, when the effective offset
charge is not at the optimal point, the three-wave mixing can also occur in
such system.

It is well known that single superconducting artificial atom can be strongly
coupled to different quantized microwave fields through the circuit QED
technique. For example, correlated emission lasing has been demonstrated using
a single three-level flux qubit circuit which is coupled to two quantized
microwave modes in a transmission line resonator, and a classical field is
coherently converted into other two different mode fields of microwave fields.
Thus, the semiclassical treatment here for microwave field can be easily
modified to the quantum case. In this case, our model can be used to study
controllable generation of single and entangled microwave photon states using
a single artificial atom. This will be very important for quantum information
processing on superconducting quantum chip.

\begin{acknowledgments}
YXL is supported by the National Basic Research Program of China Grant No.
2014CB921401, the NSFC Grants No. 61025022, and No. 91321208. Peng was
supported by ImPACT Program of Council for Science, Technology and Innovation
(Cabinet Office, Government of Japan).
\end{acknowledgments}

\appendix

\section{Parameters for driving type (1)\label{append:d12}}

\subsection{Expressions of $\bar{I}_{mn}^{\left(  1\right)  }\left(  t\right)
$}%

\begin{align}
\bar{I}_{11}^{\left(  1\right)  }\left(  t\right)   &  =I_{11}\cos^{2}%
\frac{\theta_{1}}{2}+I_{22}\sin^{2}\frac{\theta_{1}}{2}\nonumber\\
&  -I_{12}\cos\frac{\theta_{1}}{2}\sin\frac{\theta_{1}}{2}\exp\left(
-i\omega_{d1}t\right) \nonumber\\
&  -I_{21}\cos\frac{\theta_{1}}{2}\sin\frac{\theta_{1}}{2}\exp\left(
i\omega_{d1}t\right)  ,
\end{align}%
\begin{align}
\bar{I}_{22}^{\left(  1\right)  }\left(  t\right)   &  =I_{11}\sin^{2}%
\frac{\theta_{1}}{2}+I_{22}\cos^{2}\frac{\theta_{1}}{2}\nonumber\\
&  +I_{12}\cos\frac{\theta_{1}}{2}\sin\frac{\theta_{1}}{2}\exp\left(
-i\omega_{d1}t\right) \\
&  +I_{21}\cos\frac{\theta_{1}}{2}\sin\frac{\theta_{1}}{2}\exp\left(
i\omega_{d1}t\right)  ,
\end{align}%
\begin{equation}
\bar{I}_{33}^{\left(  1\right)  }\left(  t\right)  =I_{33},
\end{equation}%
\begin{align}
\bar{I}_{21}^{\left(  1\right)  }\left(  t\right)   &  =\bar{I}_{12}^{\left(
1\right)  \ast}\left(  t\right)  =\left(  I_{11}-I_{22}\right)  \cos
\frac{\theta_{1}}{2}\sin\frac{\theta_{1}}{2}\nonumber\\
&  -I_{12}\sin^{2}\frac{\theta_{1}}{2}\exp\left(  -i\omega_{d1}t\right)
\nonumber\\
&  +I_{21}\cos^{2}\frac{\theta_{1}}{2}\exp\left(  i\omega_{d1}t\right)  ,
\end{align}%
\begin{align}
\bar{I}_{31}^{\left(  1\right)  }\left(  t\right)   &  =\bar{I}_{13}^{\left(
1\right)  \ast}\left(  t\right)  =I_{31}\cos\frac{\theta_{1}}{2}\nonumber\\
&  +I_{32}\sin\frac{\theta_{1}}{2}\exp\left(  -i\omega_{d1}t\right)  ,
\end{align}%
\begin{align}
\bar{I}_{32}^{\left(  1\right)  }\left(  t\right)   &  =\bar{I}_{23}^{\left(
1\right)  \ast}\left(  t\right)  =I_{31}\sin\frac{\theta_{1}}{2}\nonumber\\
&  +I_{32}\cos\frac{\theta_{1}}{2}\exp\left(  -i\omega_{d1}t\right)  .
\end{align}

\subsection{Expressions of $K_{mk}^{\left(  1\right)  }$ and $K_{\phi
mk}^{\left(  1\right)  }$}%

\begin{equation}
K_{\phi21}^{\left(  1\right)  }=4\cos^{2}\frac{\theta_{1}}{2}\sin^{2}%
\frac{\theta_{1}}{2}\lambda_{21},
\end{equation}%
\begin{equation}
K_{\phi31}^{\left(  1\right)  }=\cos^{2}\frac{\theta_{1}}{2}\sin^{2}%
\frac{\theta_{1}}{2}\lambda_{21},
\end{equation}%
\begin{equation}
K_{\phi32}^{\left(  1\right)  }=\cos^{2}\frac{\theta_{1}}{2}\sin^{2}%
\frac{\theta_{1}}{2}\lambda_{21},
\end{equation}%
\begin{equation}
K_{21}^{\left(  1\right)  }=\cos^{4}\frac{\theta_{1}}{2}\lambda_{21},
\end{equation}%
\begin{equation}
K_{12}^{\left(  1\right)  }=\sin^{4}\frac{\theta_{1}}{2}\lambda_{21},
\end{equation}%
\begin{equation}
K_{32}^{\left(  1\right)  }=\sin^{2}\frac{\theta_{1}}{2}\lambda_{31}+\cos
^{2}\frac{\theta_{1}}{2}\lambda_{32},
\end{equation}%
\begin{equation}
K_{31}^{\left(  1\right)  }=\cos^{2}\frac{\theta_{1}}{2}\lambda_{31}+\sin
^{2}\frac{\theta_{1}}{2}\lambda_{32},
\end{equation}%
\begin{equation}
K_{23}^{\left(  1\right)  }=K_{13}=0.
\end{equation}

\section{Parameters for driving type (2)\label{append:d23}}

\subsection{Expressions of $\bar{I}_{mn}^{\left(  2\right)  }\left(  t\right)
$}%

\begin{equation}
\bar{I}_{11}^{\left(  2\right)  }\left(  t\right)  =I_{11},
\end{equation}%
\begin{align}
\bar{I}_{22}^{\left(  2\right)  }\left(  t\right)   &  =I_{22}\cos^{2}%
\frac{\theta_{2}}{2}+I_{33}\sin^{2}\frac{\theta_{2}}{2}\nonumber\\
&  -I_{23}\cos\frac{\theta_{2}}{2}\sin\frac{\theta_{2}}{2}\exp\left(
-i\omega_{d2}t\right) \nonumber\\
&  -I_{32}\cos\frac{\theta_{2}}{2}\sin\frac{\theta_{2}}{2}\exp\left(
i\omega_{d2}t\right)  ,
\end{align}%
\begin{align}
\bar{I}_{33}^{\left(  2\right)  }\left(  t\right)   &  =I_{22}\sin^{2}%
\frac{\theta_{2}}{2}+I_{33}\cos^{2}\frac{\theta_{2}}{2}\nonumber\\
&  +I_{23}\cos\frac{\theta_{2}}{2}\sin\frac{\theta_{2}}{2}\exp\left(
-i\omega_{d2}t\right) \nonumber\\
&  +I_{32}\cos\frac{\theta_{2}}{2}\sin\frac{\theta_{2}}{2}\exp\left(
i\omega_{d2}t\right)  ,
\end{align}%
\begin{align}
\bar{I}_{21}^{\left(  2\right)  }\left(  t\right)   &  =\bar{I}_{12}^{\left(
2\right)  \ast}\left(  t\right)  =I_{21}\cos\frac{\theta_{2}}{2}\nonumber\\
&  -I_{31}\sin\frac{\theta_{2}}{2}\exp\left(  i\omega_{d2}t\right)  ,
\end{align}%
\begin{align}
\bar{I}_{31}^{\left(  2\right)  }\left(  t\right)   &  =\bar{I}_{13}^{\left(
2\right)  \ast}\left(  t\right)  =I_{21}\sin\frac{\theta_{2}}{2}\nonumber\\
&  +I_{31}\cos\frac{\theta_{2}}{2}\exp\left(  i\omega_{d2}t\right)  ,
\end{align}%
\begin{align}
\bar{I}_{32}^{\left(  2\right)  }\left(  t\right)   &  =\bar{I}_{23}^{\left(
2\right)  \ast}\left(  t\right)  =\left(  I_{22}-I_{33}\right)  \cos
\frac{\theta_{2}}{2}\sin\frac{\theta_{2}}{2}\nonumber\\
&  -I_{23}\sin^{2}\frac{\theta_{2}}{2}\exp\left(  -i\omega_{d2}t\right)
\nonumber\\
&  +I_{32}\cos^{2}\frac{\theta_{2}}{2}\exp\left(  i\omega_{d2}t\right)  .
\end{align}

\subsection{Expressions of $K_{mk}^{\left(  2\right)  }$ and $K_{\phi
mk}^{\left(  2\right)  }$}%

\begin{equation}
K_{\phi21}^{\left(  2\right)  }=\cos^{2}\frac{\theta_{2}}{2}\sin^{2}%
\frac{\theta_{2}}{2}\lambda_{32},
\end{equation}%
\begin{equation}
K_{\phi31}^{\left(  2\right)  }=\cos^{2}\frac{\theta_{2}}{2}\sin^{2}%
\frac{\theta_{2}}{2}\lambda_{32},
\end{equation}%
\begin{equation}
K_{\phi32}^{\left(  2\right)  }=4\cos^{2}\frac{\theta_{2}}{2}\sin^{2}%
\frac{\theta_{2}}{2}\lambda_{32},
\end{equation}%
\begin{equation}
K_{21}^{\left(  2\right)  }=\lambda_{21}\cos^{2}\frac{\theta_{2}}{2}%
+\lambda_{31}\sin^{2}\frac{\theta_{2}}{2},
\end{equation}%
\begin{equation}
K_{32}^{\left(  2\right)  }=\lambda_{32}\cos^{4}\frac{\theta_{2}}{2},
\end{equation}%
\begin{equation}
K_{31}^{\left(  2\right)  }=\lambda_{21}\sin^{2}\frac{\theta_{2}}{2}%
+\lambda_{31}\cos^{2}\frac{\theta_{2}}{2},
\end{equation}%
\begin{equation}
K_{23}^{\left(  2\right)  }=\lambda_{32}\sin^{4}\frac{\theta_{2}}{2},
\end{equation}%
\begin{equation}
K_{13}^{\left(  2\right)  }=K_{12}^{\left(  2\right)  }=0.
\end{equation}

\section{Parameters for driving type (3)\label{append:d13}}

\subsection{Expressions of $\bar{I}_{mn}^{\left(  3\right)  }\left(  t\right)
$}%

\begin{align}
\bar{I}_{11}^{\left(  3\right)  }\left(  t\right)   &  =I_{11}\cos^{2}%
\frac{\theta_{3}}{2}+I_{33}\sin^{2}\frac{\theta_{3}}{2}\nonumber\\
&  -I_{13}\cos\frac{\theta_{3}}{2}\sin\frac{\theta_{3}}{2}\exp\left(
-i\omega_{d3}t\right) \nonumber\\
&  -I_{31}\cos\frac{\theta_{3}}{2}\sin\frac{\theta_{3}}{2}\exp\left(
i\omega_{d3}t\right)
\end{align}%
\begin{equation}
\bar{I}_{22}^{\left(  3\right)  }\left(  t\right)  =I_{22}%
\end{equation}%
\begin{align}
\bar{I}_{33}^{\left(  3\right)  }\left(  t\right)   &  =I_{11}\sin^{2}%
\frac{\theta_{3}}{2}+I_{33}\cos^{2}\frac{\theta_{3}}{2}\nonumber\\
&  +I_{13}\cos\frac{\theta_{3}}{2}\sin\frac{\theta_{3}}{2}\exp\left(
-i\omega_{d3}t\right) \nonumber\\
&  +I_{31}\cos\frac{\theta_{3}}{2}\sin\frac{\theta_{3}}{2}\exp\left(
i\omega_{d3}t\right)
\end{align}%
\begin{align}
\bar{I}_{21}^{\left(  3\right)  }\left(  t\right)   &  =\bar{I}_{12}^{\left(
3\right)  \ast}\left(  t\right)  =I_{21}\cos\frac{\theta_{3}}{2}\nonumber\\
&  -I_{23}\sin\frac{\theta_{3}}{2}\exp\left(  -i\omega_{d3}t\right)
\end{align}%
\begin{align}
\bar{I}_{31}^{\left(  3\right)  }\left(  t\right)   &  =\bar{I}_{13}^{\left(
3\right)  \ast}\left(  t\right)  =\left(  I_{11}-I_{33}\right)  \cos
\frac{\theta_{3}}{2}\sin\frac{\theta_{3}}{2}\nonumber\\
&  -I_{13}\sin^{2}\frac{\theta_{3}}{2}\exp\left(  -i\omega_{d3}t\right) \\
&  +I_{31}\cos^{2}\frac{1}{2}\theta\exp\left(  i\omega_{d3}t\right) \nonumber
\end{align}%
\begin{align}
\bar{I}_{32}^{\left(  3\right)  }\left(  t\right)   &  =\bar{I}_{23}^{\left(
3\right)  \ast}\left(  t\right)  =I_{12}\sin\frac{\theta_{3}}{2}\nonumber\\
&  +I_{32}\cos\frac{\theta_{3}}{2}\exp\left(  i\omega_{d3}t\right)
\end{align}

\subsection{Expressions of $K_{mk}^{\left(  3\right)  }$ and $K_{\phi
mk}^{\left(  3\right)  }$}%

\begin{equation}
K_{\phi21}^{\left(  3\right)  }=\cos^{2}\frac{\theta_{3}}{2}\sin^{2}%
\frac{\theta_{3}}{2}\lambda_{31},
\end{equation}%
\begin{equation}
K_{\phi31}^{\left(  3\right)  }=4\cos^{2}\frac{\theta_{3}}{2}\sin^{2}%
\frac{\theta_{3}}{2}\lambda_{31},
\end{equation}%
\begin{equation}
K_{\phi32}^{\left(  3\right)  }=\cos^{2}\frac{\theta_{3}}{2}\sin^{2}%
\frac{\theta_{3}}{2}\lambda_{31},
\end{equation}%
\begin{equation}
K_{21}^{\left(  3\right)  }=\cos^{2}\frac{\theta_{3}}{2}\lambda_{21},
\end{equation}%
\begin{equation}
K_{12}^{\left(  3\right)  }=\sin^{2}\frac{\theta_{3}}{2}\lambda_{32}%
\end{equation}%
\begin{equation}
K_{32}^{\left(  3\right)  }=\cos^{2}\frac{\theta_{3}}{2}\lambda_{32},
\end{equation}%
\begin{equation}
K_{23}^{\left(  3\right)  }=\sin^{2}\frac{\theta_{3}}{2}\lambda_{21}%
\end{equation}%
\begin{equation}
K_{31}^{\left(  3\right)  }=\cos^{4}\frac{\theta_{3}}{2}\lambda_{31},
\end{equation}%
\begin{equation}
K_{13}^{\left(  3\right)  }=\sin^{4}\frac{\theta_{3}}{2}\lambda_{31}%
\end{equation}

\end{document}